\documentclass[pra,aps,amsmath,amssymb,amsfonts,twocolumn,nofootinbib,longbibliography,superscriptaddress]{revtex4-1}
\usepackage{amssymb}
\usepackage{bm,mathrsfs}
\usepackage{graphicx}
\usepackage{epsfig}
\usepackage{amsmath,bbm}
\usepackage{amsfonts,amssymb}
\usepackage{times}
\usepackage{verbatim}
\usepackage[sort&compress]{natbib}
\usepackage{amsmath}
\usepackage{bm}
\usepackage{float}

\allowdisplaybreaks[4]
\usepackage[colorlinks,breaklinks,linkcolor=blue,anchorcolor=blue,citecolor=blue,urlcolor=blue]{hyperref}
\begin{document}
	\title{Simulation of a feedback-based algorithm for quantum optimization for a realistic neutral atom system with an optimized small-angle controlled-phase gate}
\author{S. X. Li}
\affiliation{Center for Quantum Sciences and School of Physics, Northeast Normal University, Changchun 130024,  China}

\author{W. L. Mu}	
\affiliation{Department of Physics, Beijing Normal University, Beijing, 100875, China}

\author{J. B. You}
\email{you\_jiabin@ihpc.a-star.edu.sg}
\affiliation{Institute of High Performance Computing, A*STAR (Agency for Science, Technology and Research),
1 Fusionopolis Way, Connexis, Singapore 138632}

\author{X. Q. Shao}
\email{shaoxq644@nenu.edu.cn}
\affiliation{Center for Quantum Sciences and School of Physics, Northeast Normal University, Changchun 130024, China}
\affiliation{Center for Advanced Optoelectronic Functional Materials Research, and Key Laboratory for UV Light-Emitting Materials and Technology
of Ministry of Education, Northeast Normal University, Changchun 130024, China}

\date{\today}
\begin{abstract}
In contrast to the classical optimization process required by the quantum approximate optimization algorithm, FALQON, a feedback-based algorithm for quantum optimization [A. B. Magann {\it et al.,} {\color{blue}Phys. Rev. Lett. {\bf129}, 250502 (2022)}], enables one to obtain approximate solutions to combinatorial optimization problems without any classical optimization effort. In this study, we leverage the specifications of a recent experimental platform for the neutral atom system [Z. Fu {\it et al.,} {\color{blue}Phys. Rev. A {\bf105}, 042430 (2022)}] and present a scheme to implement an optimally tuned small-angle controlled-phase gate. By examining the 2- to 4-qubit FALQON algorithms in the Max-Cut problem and considering the spontaneous emission of the neutral atomic system, we have observed that the performance of FALQON implemented with small-angle controlled-phase gates exceeds that of FALQON utilizing CZ gates. This approach has the potential to significantly simplify the logic circuit required to simulate FALQON and effectively address the Max-Cut problem, which may pave a way for the experimental implementation of near-term noisy intermediate-scale quantum algorithms with neutral-atom systems.
\end{abstract}
\maketitle
\section{Introduction}

In areas such as combinatorial optimization \cite{korte2011combinatorial} and quantum chemistry \cite{cao2019quantum}, quantum computers have the potential to outweigh their classical counterparts. The logistics and supply chain industries, as well as the pharmaceutical and biomedical research communities, can greatly benefit from combinatorial optimization \cite{drug}. Since it is NP-hard to find an exact solution to broad combinatorial optimization problems, most practical approaches focus on generating high-quality approximations instead. Finding the ground state of an Ising Hamiltonian $H_{p}$ can be seen as an analog to many combinatorial optimization problems when mapping to quantum systems \cite{Ising}. Significant progress has been made in the creation of noisy intermediate-scale quantum machines, and there is a growing interest in discovering effective algorithms meant to run on these near-term quantum devices.

Recently, the $H_p$ approximation ground state has been proposed and can be prepared using FALQON \cite{FALQON,magann2022lyapunov}, a feedback-based algorithm for quantum optimization. Unlike the quantum approximate optimization algorithm (QAOA) \cite{QAOA,9951267,PhysRevResearch.4.013141,PhysRevResearch.4.023249,PhysRevResearch.4.033029,PhysRevLett.125.260505,a12020034,Egger2021warmstartingquantum,PhysRevResearch.5.023071,TILLY20221,PRXQuantum.3.010313,guerreschi2019qaoa,9259965,fuchs2021efficient,chancellor2019domain,cerezo2021variational,harrigan2021quantum,PhysRevLett.124.090504,farhi2022quantum,lloyd2018quantum,farhi2016quantum,pagano2020quantum,PhysRevApplied.14.034009,shaydulin2019multistart,graham2022multi,wang2021noise,cerezo2021cost,stilck2021limitations,higgott2019variational,chen2020variational,khatri2019quantum,doi:10.1126/science.abo6587}, which requires a classical optimizer to calculate quantum circuit parameters, FALQON uses the output of qubit measurements to constructively assign values to quantum circuit parameters. Consequently, FALQON overcomes the challenge of optimizing a large number of variational parameters, which is a significant obstacle to QAOA scalability. However, the downside is that this requires numerous layers in the quantum circuit, leading to an accumulation of quantum logic gate errors as the number of layers increases. Therefore, the focus shifts to reducing the total number of two-qubit entangling gates and improving their accuracy, given the successful realization of high-fidelity single-qubit gates.

Two-qubit entanglement gates, such as controlled-phase gates, play a central role in universal quantum computing \cite{galindo2002information,ladd2010quantum,wendin2017quantum}. Experimentally, the achievement of a fast and high-fidelity controlled-phase gate has been demonstrated in various physical systems, including nuclear magnetic resonance, quantum dots, ion traps, semiconductor silicon, and Josephson junctions \cite{calderon2019fast,kanaar2021single,ballance2016high,gaebler2016high,barends2014superconducting,wang2019experimental,rol2019fast}. Nevertheless, researchers have predominantly focused on realizing controlled-$Z$ gates and equivalent controlled-NOT gates in universal quantum computing schemes, with little attention paid to the implementation of controlled-phase gates with small angles.
In practical applications, the small-angle controlled-phase gate is often dismissed as mathematically approximate to the unit matrix, leading to doubts about its practical significance. However, when combining the insights from the literature \cite{improve}, using small-angle controlled-phase gates to decompose the phase-separation unitary operator can significantly reduce the number of two-qubit entangling gates and improve the performance of FALQON. Consequently, the realization of small-angle controlled-phase gates in real physical systems has become a subject worth investigating in the context of FALQON implementation.

Due to stable encoding in hyperfine atomic states and the ability to manipulate and measure qubit states with laser light \cite{bloch2008quantum,QuantumInformationWithRydbergAtoms,browaeys2016experimental,shao2024rydberg}, neutral atoms have emerged as a promising physical system for quantum information processing. Recently, researchers have utilized Rydberg atom arrays for quantum optimization to solve the Maximum Independent Set problem on unit disk graphs \cite{doi:10.1126/science.abo6587}. However, this method lacks universality since the researchers organically integrated the blockade of Rydberg atoms with the characteristics of this problem, which may not be applicable to other combinatorial optimization problems. A more general approach would be to use Rydberg atoms to implement quantum logic gates in quantum circuits. The strong interaction between Rydberg atoms \cite{urban2009observation,tong2004local,gaetan2009observation} has led to the theoretical proposal \cite{jaksch2000fast,lukin2001dipole,brion2007implementing,wu2010implementation,muller2011optimizing,xia2013analysis,petrosyan2014binding,sarkany2015long,saffman2016quantum,su2016one,su2017fast,shi2018deutsch,shi2017annulled,shi2017rydberg,huang2018robust,su2018one,li2018unconventional,shi2019fast,yin2020one,li2022optimal,li2022coherent} and the successful experimental implementation of quantum logic gates \cite{levine2019parallel,martin2021molmer,fu2022high}. Recently, our group proposed a method to realize the continuous controlled-phase gate set in the Rydberg blockade mechanism, based on adiabatic evolution by considering a symmetric two-photon excitation process via the second resonance line \cite{li2022single}. Through a single-temporal modulation coupling of the ground state and the intermediate state, the logical qubit state $|11\rangle$ alone can accumulate a dynamic phase factor for the controlled-phase gate, adjustable from 0.08$\pi$ to $\pi$ by tuning the shape of the temporal pulse.
Still, some challenges remain in this scheme. First, from an experimental standpoint, it requires a high-power laser to couple the excitation of the intermediate state to the Rydberg state, with a Rabi frequency of approximately 200 MHz, limiting its implementation on various neutral-atom experimental platforms. Second, the scheme approximates the controlled-phase gate of small angle with a unit operator, making it unsuitable for the implementation of the FALQON algorithm.

\begin{figure}
\includegraphics[width=0.9\linewidth]{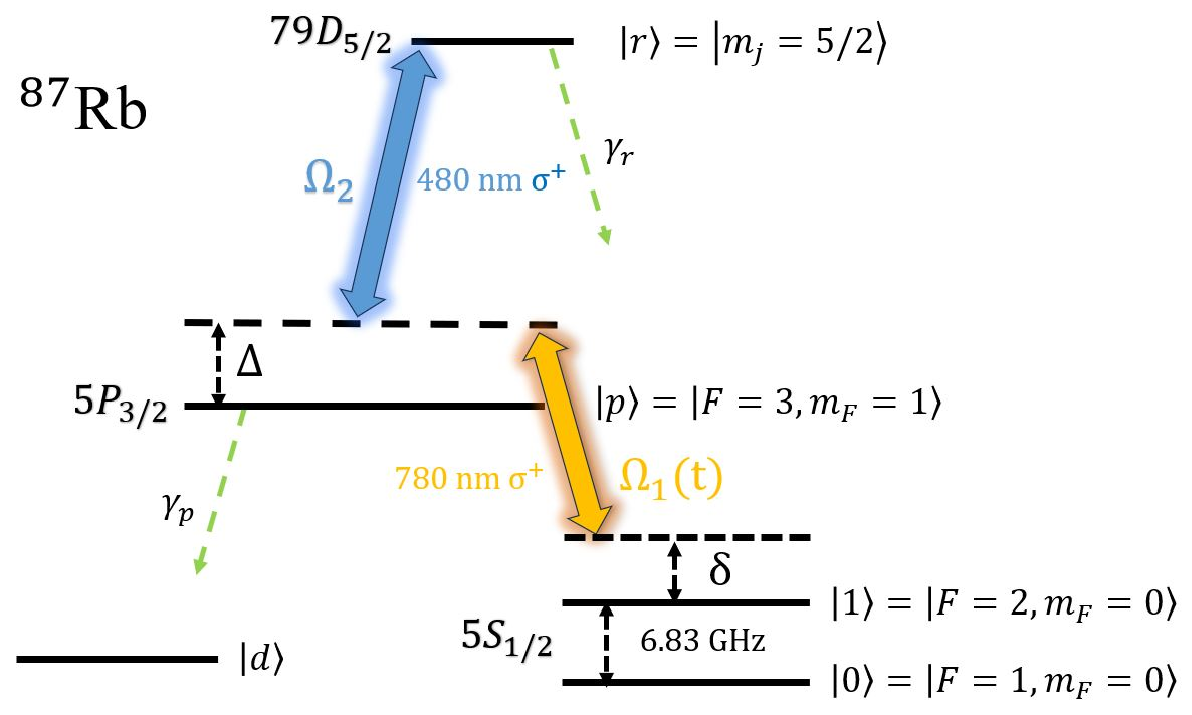}
\caption{Relevant configuration of level for $^{87}$Rb atom used in Ref.~\cite{fu2022high}.}\label{fig1}
\end{figure}

In this work, we focus on exploring how to use the experimental platform introduced in~\cite{fu2022high} to achieve robust small-angle controlled-phase gates and utilize them to simulate the FALQON algorithm.
We continue our study of a native Rydberg two-photon excitation mechanism similar to that described in Ref.~\cite{li2022single}, as shown in Fig.~\ref{fig1}, where the transition from the intermediate state to the Rydberg state is driven by a laser with a Rabi frequency of up to 50~MHz. Meanwhile, the transition from the ground state to the intermediate state is facilitated by a time-dependent Gaussian pulse under optimal parameters. Unlike the model presented in Ref.~\cite{li2022single}, which only considers the two-photon resonance, we have found in our subsequent discussions on the fluctuation of two-photon detuning that taking into account an appropriate two-photon detuning $\delta$ can enhance the fidelity of small-angle controlled-phase gates. As a result, our scheme can achieve a fast, high-fidelity, and resilient small-angle controlled-phase gate and enhances the performance of FALQON for a realistic neutral-atom system.

The remainder of the paper unfolds as follows. We begin by providing a concise overview of the FALQON principle in Sec.~\ref{sec2}. Next, in Sec.~\ref{sec3}, we elaborate on enhancing FALQON's performance with 
a realistic physical system. In Section~\ref{sec4}, we demonstrate the efficacy of FALQON using the controlled-phase gate with a small angle, by analyzing in detail 2 to 4 qubit Max-Cut problems. Additionally, we delve into practical considerations such as experimental feasibility and the impact of technical imperfections, such as fluctuations of the external field and relevant noise. Finally, our findings are summarized in Sec.~\ref{sec5}.

\section{Brief review of FALQON}\label{sec2}

Many discrete combinatorial optimization problems can be encoded into an Ising
Hamiltonian $H_{p}$ (problem Hamiltonian) \cite{Ising}:
\begin{equation}\label{eq1}
H_{p}=\sum_{i\textless j}J_{ij}Z_{i}Z_{j}+\sum_{i=1}^n h_{i}Z_{i},
\end{equation}
where $Z_{i}$ is the Pauli-$Z$ operator for $i$-th qubit. The problem Hamiltonian corresponds to the objective function of the given combinatorial optimization problem, whose minimum represents the presence of solution. Therefore, the problem of finding the solution that minimizes the objective function can be converted into finding the ground state $|\psi\rangle$ of $H_p$ \cite{QAOA}.

The Max-Cut problem is one of the most well-known optimization problems, attempting to partition the vertices of
a graph into two sets so that the maximum number of edges can be cut. As a typical NP-hard problem, its problem Hamiltonian is $H_{p}=1/2 \sum^{m}_{\alpha=1}(Z_{\alpha_1}Z_{\alpha_2}-1)$, in which $\alpha_{1,2}$ are qubit indices representing the vertices of the edge $\alpha$. The standard approach in the quantum circuit-based QAOA to find the ground state of $H_{p}$ involves measuring the probability distribution of the final state in the $l$-th layer, which can be represented as $|\gamma \beta\rangle=e^{-i\beta_{l} H_{d}}e^{-i\gamma_{l}H_{p}}......e^{-i\beta_{1} H_{d}}e^{-i\gamma_{1}H_{p}}|\psi_{0}\rangle$ \cite{QAOA}. The optimal values for all $\gamma$ and $\beta$ can be determined using classical optimization techniques. In the limit as $l$ approaches infinity, the above result can be interpreted as a trotterized version of the adiabatic evolution from the initial state to the ground state of $H_p$.

In contrast to QAOA, FALQON does not require any classical optimization efforts and can still help to find the approximate ground state of $H_p$ \cite{FALQON,magann2022lyapunov}. Its core idea is to achieve a monotonic decrease of $\left\langle H_{p}\right\rangle$ over layers.
As the number of layers increases, the $l$th layer of $\left\langle H_{p}\right\rangle$ will become closer and closer to $ \left\langle H_{p}\right\rangle _{\rm min}$ and thus an approximate ground state is prepared.

To start with, consider a quantum system whose dynamic is governed by
\begin{equation}\label{eq2}
i\frac{d}{dt}|\psi(t)\rangle=[H_{p}+H_{d}\beta(t)]|\psi(t)\rangle,
\end{equation}
where $H_{d}$ is control Hamiltonian, and $\beta(t)$ is a time-dependent control function to the system. In order to make $ \left\langle H_{p}\right\rangle$ monotonic decrease over time, $\beta(t)$ can be set as
\begin{equation}\label{eq4}
\beta(t)=-kA(t),
\end{equation}
with $k>0$ represents the strength of the feedback signal (In this work, we consider a fixed value of $k=2$, which is enough in practice) to make
\begin{equation}\label{eq3}
\frac{d}{dt}\left\langle H_{p}\right\rangle=A(t)\beta(t)\leq 0,
\end{equation}
where $A(t)\equiv \langle\psi(t)|i[H_{d},H_{p}]|\psi(t)\rangle$.

Suppose that the time interval $2\Delta t$ of the evolution of each layer is small enough and that the continuous-time evolution of the system can be approximated by trotterization. After evolution, the approximate ground state of a $l$-layer FALQON is prepared \cite{FALQON,magann2022lyapunov}:
\begin{equation}\label{eq5}
|\psi_{l}\rangle=e^{-i\beta_{l} \Delta tH_{d}}e^{-i\Delta tH_{p}}......e^{-i\beta_{1} \Delta tH_{d}}e^{-i\Delta tH_{p}}|\psi_{0}\rangle,
\end{equation}
where  $H_{d}=-1/2\sum_{i}X_{i}$ with $X$ being the Pauli-$X$ operator, and $|\psi_{0}\rangle$ is a ground state of $H_{d}$ that can be easily initialized experimentally, and $\beta_j$ of $j$-th layer is determined by the feedback signal of $(j-1)$-th layer without the need for an expensive classical optimization loop, i.e.,
\begin{equation}\label{eq6}
\beta_{j}=-k\langle\psi_{j-1}|i[H_{d},H_{p}]|\psi_{j-1}\rangle.
\end{equation} 

The evaluation of FALQON performance is achieved using two significant metrics: the approximation ratio $r_{A}=\left\langle H_{p}\right\rangle/\left\langle H_{p}\right\rangle _{\rm min}$ and the success probability $\phi$ associated with the measurement of ground states. To optimize the efficiency of FALQON when implementing it in a practical quantum system, it is advantageous to consider the requisite two-qubit entangling gates from two distinct vantage points: quantity and quality.

\section{route to improving the performance of FALQON}\label{sec3}
 \subsection{Quantity: reducing the number of two-qubit entangling
gates}\label{subsec2.2}

As the errors in quantum circuit implementation are primarily attributed to two-qubit entangling gates, we assume that all single-qubit gates are ideal and focus on reducing the number of two-qubit entangling gates.

From an algorithm perspective, the number of two-qubit entangling gates scales with the layers. To address this, one possible approach is to reduce the number of layers without compromising FALQON's quality.

A rational increase in $\Delta t$ may be effective in reducing the number of layers. Although a smaller $\Delta t$ can theoretically improve the quality of the trotterized approximation applied to Eq.~(\ref{eq2}) and enhance the final convergent value of $r_{A}$, it leads to an increase in layers, which is detrimental to the practical implementation of FALQON. On the other hand, as long as the condition in Eq.~(\ref{eq3}) is satisfied, a larger $\Delta t$ can accelerate the convergence of $r_{A}$ and potentially reduce the number of layers while maintaining the performance of FALQON. In most situations, the critical value of $\Delta t$ decreases with an increase in the size of the combinatorial optimization problem, and further details can be found in Refs. \cite{FALQON,magann2022lyapunov}. Therefore, the choice of rational $\Delta t$ plays a crucial role in enhancing the experimental performance of FALQON.

From the implementation perspective of the quantum circuit, another possible approach to reduce the number of two-qubit entangling gates is to rationally decompose the unitary operator of the $ZZ$ term in the problem Hamiltonian from Eq.~(\ref{eq1}).

\begin{figure}
\includegraphics[width=1\linewidth]{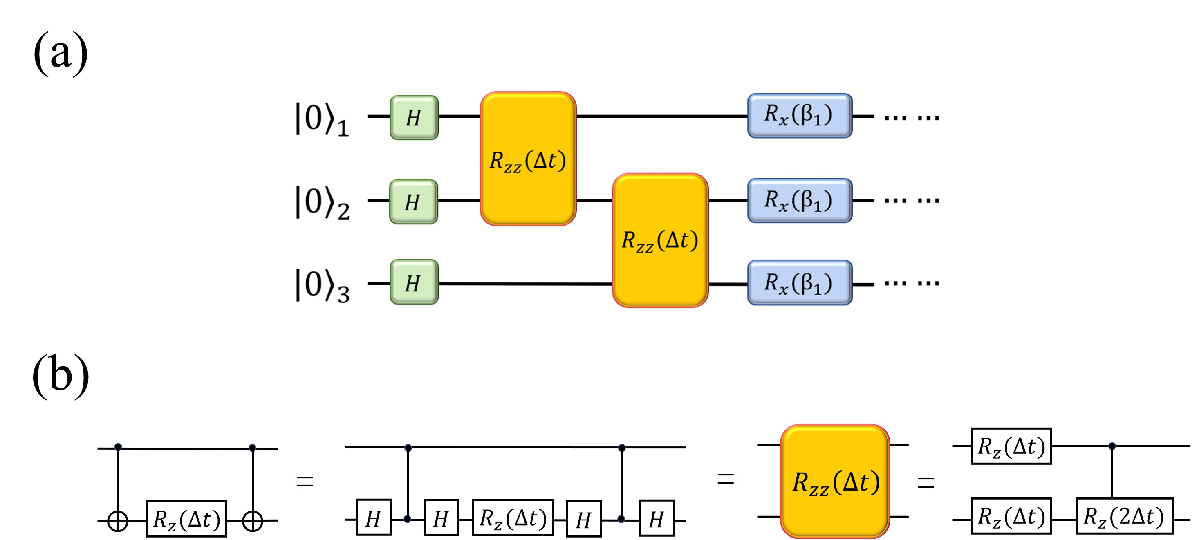}
\caption{(a) Circuit diagram for 3-qubit FALQON only with ZZ interaction term $Z\otimes Z/2$ in problem Hamiltonian showing a single-layer cycle. (b) Decomposition of the
$R_{zz}(\Delta t) = e^{-i\Delta t Z\otimes Z/2}$ interaction.}\label{figx}
\end{figure}

The quantum circuit used for the implementation of FALQON is shown in Fig.~\ref{figx}(a), without explicitly depicting the single qubit operators of $H_p$.
In implementing FALQON, there are several approaches to decompose the unitary operator of the $ZZ$ term in Eq.~(\ref{eq1}). As shown in the left side of Fig.~\ref{figx}(b), a common method is to decompose the phase-separation unitary operators into a gate sequence consisting of two CZ gates, combined with several single-qubit gates \cite{otterbach2017unsupervised,bengtsson2019quantum}, or two CNOT gates, combined with a $R_{z}$ gate. This approach requires two two-qubit entangling gates per phase-separation unitary operator. To reduce the number of two-qubit entangling
gates, it is advantageous to use mall-angle controlled-phase gates combined with several single-qubit gates to decompose the phase-separation unitary operators as shown in the right side of Fig.~\ref{figx}(b) when implementing FALQON. In this approach, only one two-qubit entangling gate is needed per phase-separation unitary operator, thus reducing errors caused by such gates \cite{improve,li2022single}.

 \subsection{Quality: optimizing the small-angle controlled-phase gate}\label{subsec2.3}
To evaluate the feasibility of our approach, we will use the experimental parameters described in Ref.~\cite{fu2022high} to assess the effectiveness of the implemented small-angle controlled-phase gate.
 The configuration of the atomic levels is depicted in Fig.~\ref{fig1}(a). The logic qubits are encoded
into the hyperfine ground state of the $^{87}$Rb atom: $|{0}\rangle=|5S_{1/2}, F=1,m_{F}=0\rangle$ and $|{1}\rangle=|5S_{1/2}, F=2,m_{F}=0\rangle$. Additionally, we have introduced an uncoupled state $|{d}\rangle$ to represent the leakage level outside $|{0}\rangle$ and $|{1}\rangle$ for simplicity. The interaction between the atoms is mediated by the Rydberg state $|{r}\rangle=|{79D_{5/2},m_{j}=5/2}\rangle$. To coherently drive the atoms from the ground states to the Rydberg states, we utilize two-photon excitation lasers with counterpropagating $\sigma^{+}$-polarized 780-nm  laser and $\sigma^{+}$-polarized  480-nm laser. Under the premise of considering spontaneous radiation,
the master equation of the system
in Lindblad form reads
\begin{equation}\label{eq10}
\dot{\rho}= -i[H(t),\rho]+\mathcal{L}_p[\rho]+\mathcal{L}_r[\rho],
\end{equation}
where
\begin{eqnarray}\label{eq8}
H(t)&=&\sum_{i=a,b} \frac{1}{2}\Omega_1(t)|p_i\rangle\langle1_i|+\frac{1}{2}\Omega_2|r_i\rangle\langle p_i|+\textup{H.c.}\nonumber\\&&-\Delta|p_i\rangle\langle p_i|-\delta|{1}_i\rangle\langle{1}_i|
+u_{rr}|rr\rangle\langle rr|,
\end{eqnarray}
describes the coherent dynamics of the system
and
\begin{equation}\label{eq11}
\mathcal{L}_p[\rho]=\sum_{i=a,b}\sum_{\mu=0,1,d}L_{\mu p}^{(i)}\rho L_{\mu p}^{(i)\dagger}-\frac{1}{2}\left\{L_{\mu p}^{(i)\dagger}L_{\mu p}^{(i)},\rho\right\},
\end{equation}

\begin{equation}\label{eq12}
\mathcal{L}_r[\rho]=\sum_{i=a,b}\sum_{\nu=0,1,d,p}L_{\nu r}^{(i)}\rho L_{\nu r}^{(i)\dagger}-\frac{1}{2}\left\{L_{\nu r}^{(i)\dagger}L_{\nu r}^{(i)},\rho\right\},
\end{equation}
picture the spontaneous emission from the intermediate state $|p\rangle$ and Rydberg state $|r\rangle$ to the ground states.
The jump operator $L_{\mu p(\nu r)}^{(i)}=\sqrt{b_{\mu p(\nu r)}\gamma_{p(r)}}|\mu(\nu)_{i}\rangle\langle p(r)_i|$, where the branching ratios are $b_{0(1)p}=1/8$, $b_{0(1)p}=1/8$, $b_{dp}=3/4$, $b_{1(0)r}=1/16$, $b_{dr}=3/8$, and $b_{pr}=1/2$. At room temperature
(300~K), the lifetime of state $|p\rangle$ and $|r\rangle$ are $\tau_{p}=1/\gamma_p=0.0262$~$\mu$s and $\tau_{r}=1/\gamma_r=212$~$\mu$s.
The atomic separation is established at a value of 3.6~$\mu$m, while the associated Rydberg interaction strength is measured to be $u_{rr}/2\pi=1.855$~GHz \cite{vsibalic2017arc}.

In the conducted experiment \cite{fu2022high}, it was theoretically determined that the Rabi frequency $\Omega_2$ of 480-nm laser light can reach a value of 91~MHz. This calculation was based on the parameters provided with a power of 120~mW and a beam waist of 8.3~$\mu$m. However, due to the optical route loss experienced during the experiment, the available Rabi frequency is limited to 50~MHz. The time-dependent Rabi frequency $\Omega_1(t)$ of the 780~nm laser can be generated by commercially available arbitrary waveform generators. $\Delta$ and $\delta$ represent single-photon detuning and two-photon detuning, respectively.

The fidelity $F$ of the gate can be defined as
\begin{equation}
F=\rm{tr}[\rho_t\rho_{\rm{target}}],
\end{equation}
where $\rho_t$ represents the state density matrix after the evolution governed by Eq.~(\ref{eq10}) from the initial state $(|00\rangle+|01\rangle+|10\rangle+|11\rangle)/2$, and $\rho_{\rm{target}}$ represents the ideal density matrix of target state $(|00\rangle+|01\rangle+|10\rangle+e^{-i\theta}|11\rangle)/2$.

\begin{figure}
\includegraphics[width=1\linewidth]{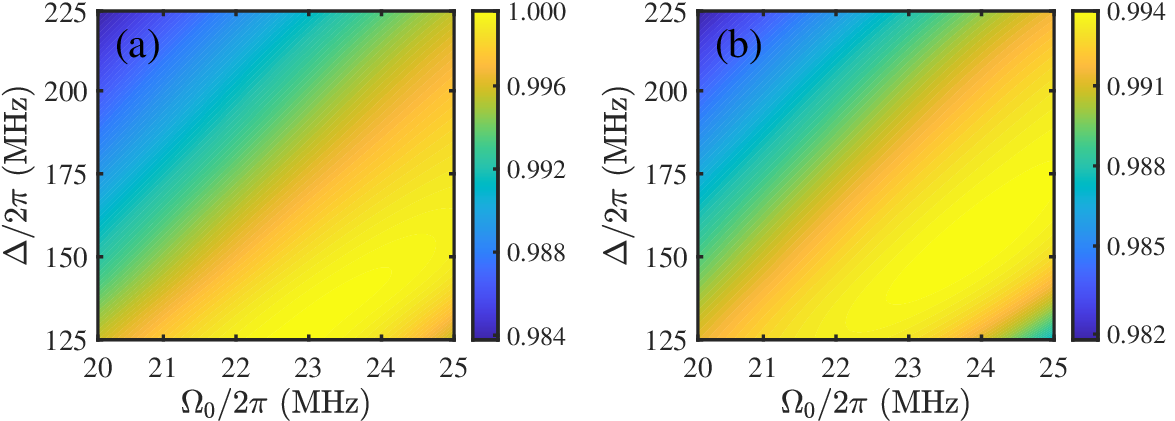}
\caption{The fidelity of the controlled-phase gate of the Rydberg atom system controlled by the Gaussian pulse $\Omega_0 \exp[-(t-2T)^2/T^2]$, which can pick up a phase $\theta=2\Delta t=0.4$~rad on the $|11\rangle$, is as a function of variational parameters ($\Omega_0$, $\Delta$) (a) via
Schr\"{o}dinger equation simulation and (b) via master equation simulation ($\delta /2\pi=0$~kHz). 
}\label{fig2}
\end{figure}

\begin{figure*}
\centering
\includegraphics[width=1\linewidth]{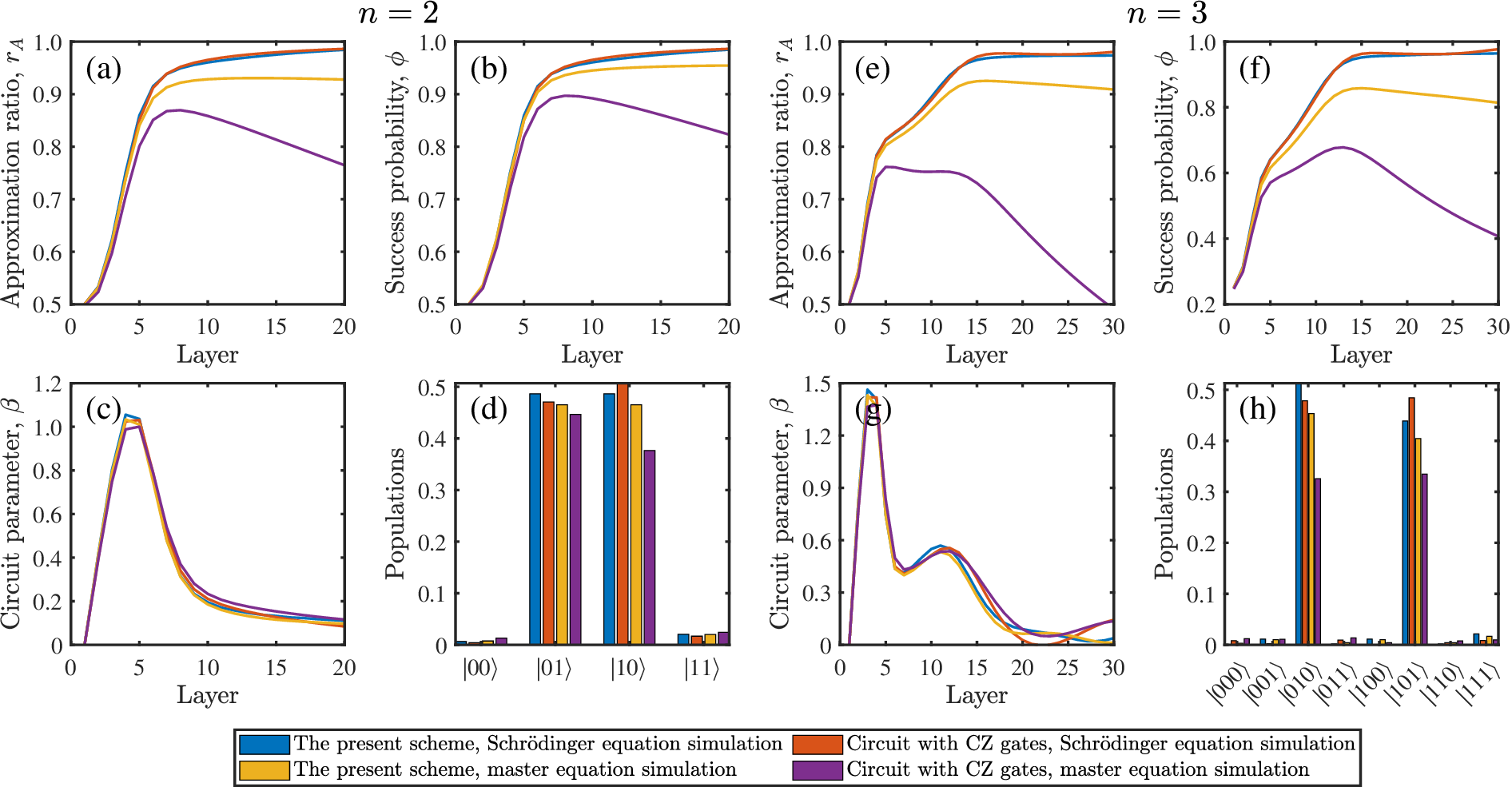}
\caption{The numerical simulation results of 2-qubit and 3-qubit FALQON  implemented by the controlled-phase gates via Gaussian pulse with phase $\theta=0.4$~rad ($\Delta t=0.2$, $\delta /2\pi=0$~kHz). The approximation ratio $r_{A}$ of each layer is plotted in (a),(e). The success probability $\phi$ is plotted in (b),(f). The values of $\beta$ are plotted in (c),(g). The measure probabilities of logical states at the 14th and 15th layer are plotted in (d),(h), respectively.}\label{fig3}
\end{figure*}

In the foundational literature on FALQON, Ref.~\cite{FALQON}, the authors delved into a detailed discussion of a 3-qubit Max-Cut instance, specifically selecting $\Delta t=0.2$. For example, they demonstrated that, through ideal simulation, FALQON can ensure a consistent increase in the energy approximation ratio $r_{A}$ as the number of layers increases. In this article, we first examine the implementation of this Max-Cut instance using our approach, and we also choose $\Delta t=0.2$. Hence, we need to implement controlled-phase gates with a phase of $\theta=2\Delta t=0.4$~rad in quantum circuits. First, we consider the case of a two-photon resonance. In Fig.~\ref{fig1}, we initially demonstrate the fidelity of the controlled-phase gate in relation to the variational parameters ($\Omega_0$, $\Delta$) when the system is driven by a Gaussian pulse $\Omega_0\exp[-(t-2T)^2/T^2]$, following the procedure to implement the parameterized controlled phase gate as outlined in \cite{li2022single}. To achieve a rapid and high-fidelity logic gate, we constrain the gate evolution time to be 1~$\mu$s, which implies that the duration of the Gaussian pulse, indicated as $4T$, is set to 1~$\mu$s. In this context, to identify the optimal parameters for fidelity, we extensively investigated the impact of parameters ($\Omega_0$, $\Delta$) on fidelity across a wide range. The main high-fidelity region, exceeding 0.98, is illustrated in Fig.~\ref{fig2}. When spontaneous emission is not taken into account, the fidelity of the controlled-phase gate can exceed 0.9991 by appropriately selecting parameters. However, when accounting for spontaneous emission, the maximum fidelity of the controlled-phase gate with a phase of $\theta=0.4$~rad is reduced to 0.9940 (rounded to 4 significant figures) by configuring $\left\{\Omega_0/2\pi=24,\Delta/2\pi=160\right\}$~MHz.

\section{numerical results}\label{sec4}

In our discussion, we explore the role of small-angle controlled-phase gates in the FALQON framework to tackle the Max-Cut problem and compare it to an alternative method to implement FALQON, which uses CZ gates on the same quantum hardware. In Ref.~\cite{fu2022high}, the authors presented a protocol to achieve high-fidelity CZ gates, along with detailed parameter settings. Taking into account a graph composed of $n=2$ nodes linked by a single edge, the corresponding problem Hamiltonian is represented as $H_{p}=1/2(Z_{1}Z_{2}-1)$. Extending to an unweighted graph with $n=3$ nodes interconnected by two edges, the problem Hamiltonian becomes $H_{p}=1/2(Z_{1}Z_{2}-1)+1/2(Z_{2}Z_{3}-1)$. In implementation, we consider a 20-layer evolution for the 2-qubit FALQON and a 30-layer evolution for the 3-qubit FALQON. The approximation ratio $r_{A}$ of FALQON from our numerical simulations is illustrated in Figs.~\ref{fig3}(a) and ~\ref{fig3}(e). As shown by the blue and orange lines, both schemes exhibit nearly equivalent performance, and FALQON exhibits a consistent increase in $r_{A}$ in the 2 and 3 qubit cases examined without accounting for spontaneous emission. When spontaneous emission is taken into account, the performance of FALQON in both schemes decreases, as indicated in Figs.~\ref{fig3}(a) and ~\ref{fig3}(e). 
In the scheme utilizing CZ gates, despite the high fidelity of individual CZ gates, the number of two-qubit entangling gates required for the same number of layers is doubled. Consequently, the impact of spontaneous emission on the scheme implemented with CZ gates is significantly greater, as illustrated by the purple lines in Figs.~\ref{fig3}(a) and ~\ref{fig3}(e). Another aspect of assessing the performance of FALQON involves the probability of success $\phi$. The numerical simulation results for $\phi$ are displayed in Figs.~\ref{fig3}(b) and ~\ref{fig3}(f). In the 2-qubit instance examined, $\phi$ represents the combined measurement probabilities of the ground states $|01\rangle$ and $|10\rangle$. Moreover, for the 3-qubit instance under consideration, the ground states are $|010\rangle$ and $|101\rangle$. The lines in Figs.~\ref{fig3}(c) and ~\ref{fig3}(g) illustrate the values of $\beta$ determined by the feedback data. In the presence of spontaneous emission, the success probability $\phi$ peaks at layer 14 for the 2-qubit FALQON and at layer 15 for the 3-qubit FALQON, as shown as the yellow lines in Figs.~\ref{fig3}(b) and ~\ref{fig3}(f). At these layers, the measurement probabilities for various states of logical computational basis are depicted by the yellow bars in Figs.~\ref{fig3}(d) and ~\ref{fig3}(h). In contrast to the blue bars representing simulations without spontaneous emission, these results underscore the challenges posed by spontaneous emission for these problem instances. Excluding single-qubit gates, the total evolution times for circuits adopting CZ gates are 80~$\mu$s ($n=2$) and 240~$\mu$s ($n=3$), whereas for circuits employing small-angle controlled-phase gates, the overall evolution times are reduced to 20~$\mu$s ($n=2$) and 60~$\mu$s ($n=3$). For more complex systems, the need for additional layers to approach the solution increases, leading to an accumulation of two-qubit gate errors. This underscores the advantages of the FALQON scheme using small-angle controlled-phase gates: it not only enhances algorithm performance when accounting for spontaneous emission, but also reduces execution time.

\begin{figure}
\includegraphics[width=1\linewidth]{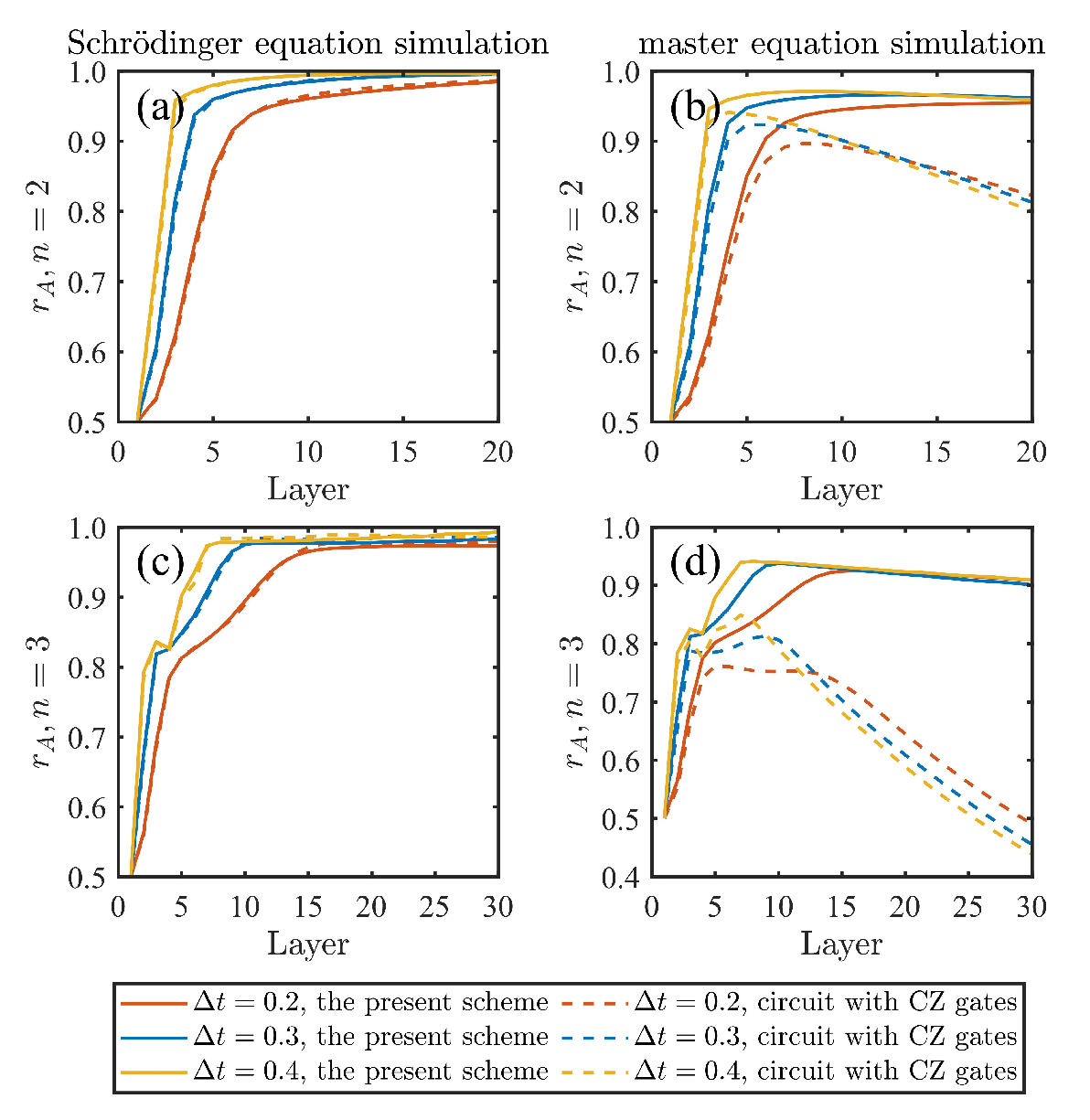}
\caption{The numerical simulation results of $r_{A}$ in (a),(b) 2-qubit and (c),(d) 3-qubit FALQON when selecting different $\Delta t$. Among these, (a),(c) are simulated by Schr\"{o}dinger equation, and (b),(d) are simulated by master equation. The solid lines represent the scheme implemented with small-angle controlled-phase gates, and the dashed lines represent the scheme implemented with CZ gates ($\delta /2\pi=0$~kHz).}\label{fig4}
\end{figure}

Now, we study the influence of $\Delta t$ on the performance of FALQON in the physical system considered. By judiciously increasing the value of $\Delta t$ without violating the condition in Eq.~(\ref{eq3}), the performance of FALQON can be somewhat improved, especially when accounting for spontaneous emission. As shown in Fig.~\ref{fig4}, rational selection of a larger $\Delta t$ can accelerate the convergence of $r_{A}$. When ignoring the effects of spontaneous emission and selecting larger layers $\Delta t$, the layers needed by FALQON to achieve the same performance can be reduced in both schemes. This method will continue to exhibit varying degrees of improvement for both schemes, even when considering spontaneous emission. Therefore, within the same combinatorial optimization problem, rational increase $\Delta t$ can be a good way to reduce the number of layers and improve the experimental feasibility.

\begin{figure}
\includegraphics[width=1\linewidth]{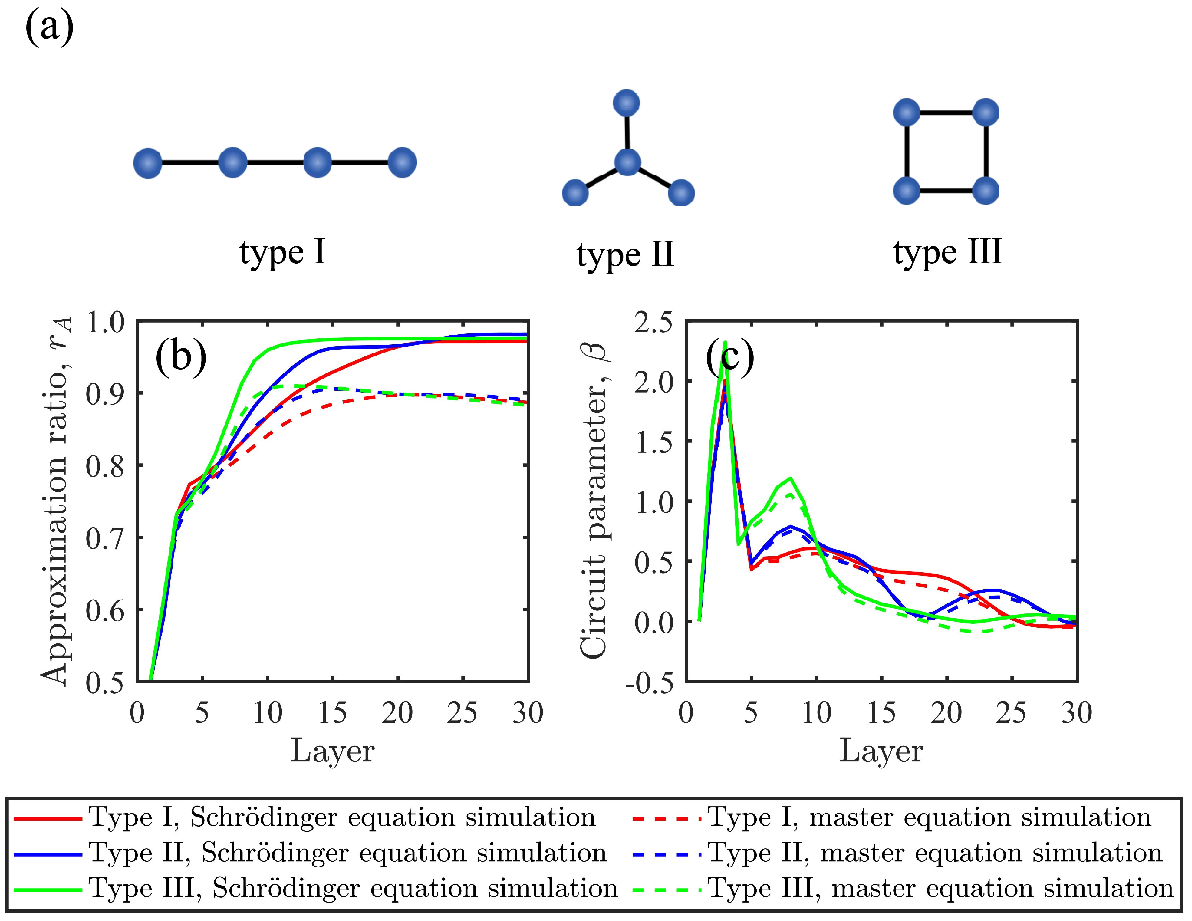}
\caption{The numerical simulation results of 4-qubit FALQON. (a) The considered three different types of graphs in 4-qubit Max-Cut problem. (b) The approximation ratio $r_{A}$ of each layer. (c) The circuit parameter $\beta$ of each layer. The solid lines are simulated by Schr\"{o}dinger equation, and the dashed lines are simulated by master equation ($\delta /2\pi=0$~kHz).}\label{fig5}
\end{figure}

When the scale of the Max-Cut problem is extended to involve more qubits, the number of different graph types also increases correspondingly. Therefore, in this study, we scale up the Max-Cut problem to a four-qubit scenario and investigate the impact of different graph types on the FALQON algorithm by considering three common types of graph of Fig.~\ref{fig5}(a). The results of the numerical simulation of the approximation ratio $r_{A}$ are depicted in Fig.~\ref{fig5}(b). It is obvious from the outset that when expanded to encompass four qubits, the FALQON algorithm remains effective, achieving high levels of $r_{A}$ regardless of whether spontaneous emission is considered or not. Among the considered cases, while the type of graph does not appear to affect the final converged value very much, it does have an influence on the rate of convergence. Within the same interval $\Delta t$, the graph type \uppercase\expandafter{\romannumeral3} with the best symmetry converges the fastest, followed by the type \uppercase\expandafter{\romannumeral2}, with the type \uppercase\expandafter{\romannumeral1} showing the slowest convergence rate. The detailed parameter $\beta$ of each layer is shown in Fig.~\ref{fig5}(c). This may imply that when a fixed $\Delta t$ is set to handle the Max-Cut problem of graphs of different types but with the same number of qubits, the FALQON algorithm may perform better on graphs with better symmetry.

\begin{figure}
\includegraphics[width=1\linewidth]{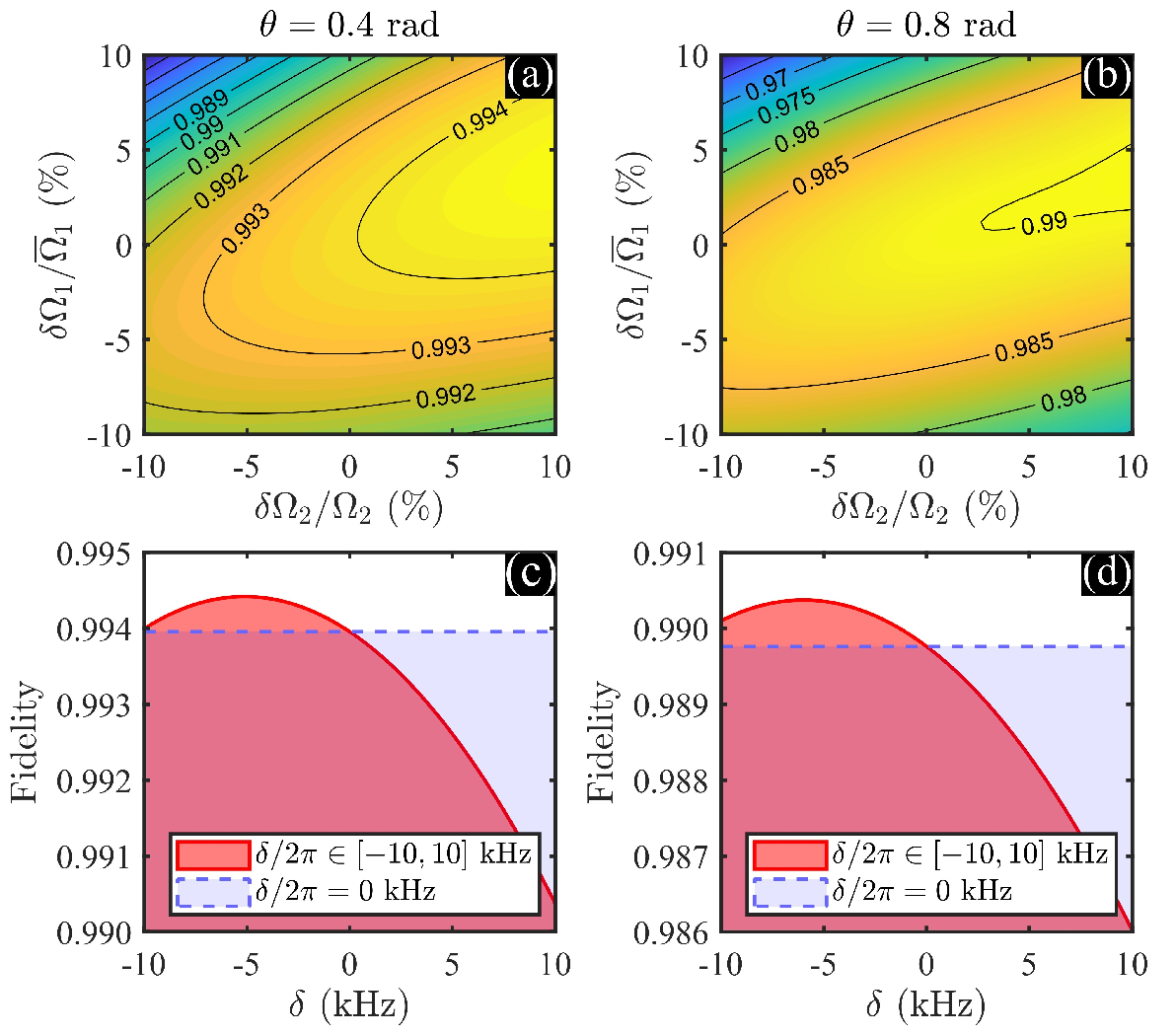}
\caption{The fidelity of the optimally-tuned controlled-phase gate, with angles (a),(c) $\theta=0.4$~rad and (b),(d) $\theta=0.8$~rad, is assessed under two distinct conditions: (a),(b) laser intensity fluctuation, and (c),(d) two-photon detuning fluctuation.}\label{fig6}
\end{figure}

\begin{figure}
\includegraphics[width=1\linewidth]{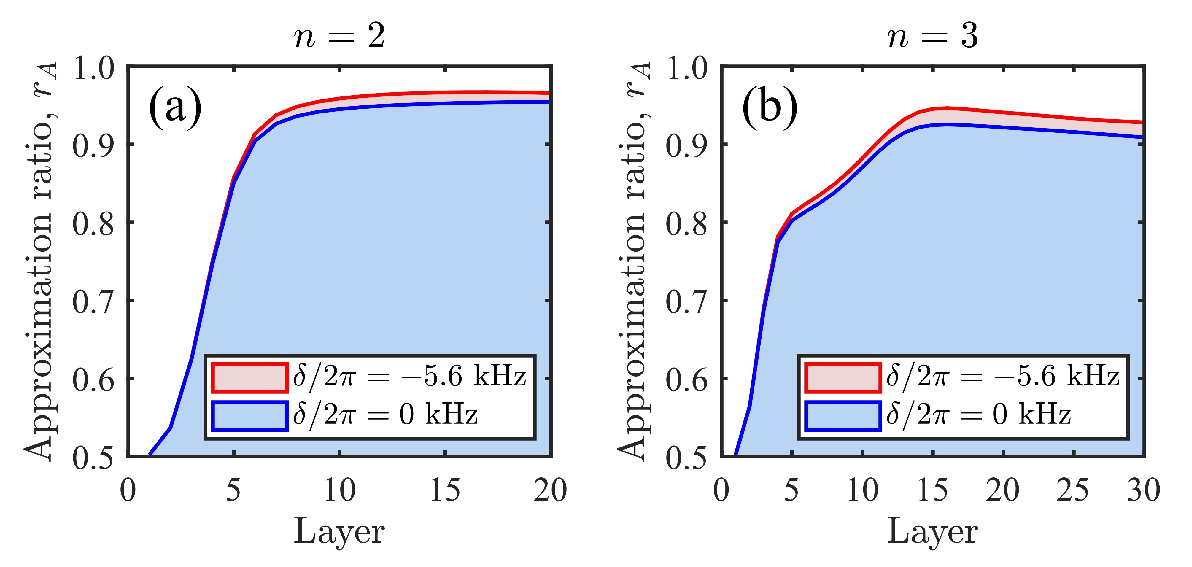}
\caption{The numerical master equation simulation results of $r_{A}$ of (a) 2-qubit FALQON and (b) 3-qubit FALQON ($\Delta t=0.2$) under $\delta /2\pi=-5.6$~kHz (red lines) and $\delta /2\pi=0$~kHz (blue lines) implemented with small-angle controlled-phase gates.}\label{fig7}
\end{figure}

In the preceding discussions, the logic gate error was only considered for spontaneous emission, which is intrinsic to the system and impossible to avoid. When multiple fields are applied in the experiments of neutral-atom systems, the intensity fluctuation of laser fields and the two-photon $\delta$ fluctuation will also have an impact on the fidelity of logic gates. The system Hamiltonian considering the intensity fluctuation of laser fields and the two-photon $\delta$ fluctuation can be written as
\begin{eqnarray}\label{eq17}
H_{\Omega,\delta}&=&\sum_{i=a,b} \frac{1}{2}(\Omega_{1}(t)+\delta \Omega_{1})|{p}_i\rangle\langle{1}_i|+\frac{1}{2}(\Omega_{2}+\delta \Omega_{2})|{r}_i\rangle\langle{p}_i|\nonumber\\&&+\textup{H.c.}-\Delta|{p}_i\rangle\langle{p}_i|
-\delta|{1}_i\rangle\langle{1}_i|+u_{rr}|rr\rangle\langle rr|,
\end{eqnarray}
where $\delta \Omega_{i}$ represents the time-dependent fluctuation introduced on the driving Rabi frequency. The numerical results of these forms of noise that affect the fidelity of logic gates can be individually shown in Fig.~\ref{fig6}, where $\overline{\Omega}_{1}=\int \Omega_1(t)dt/4T$. This analysis of fluctuations indicates that fluctuations do not necessarily have a negative impact on the fidelity of logic gates. This has inspired us to start with a higher value of $\Omega_2$, and an appropriate value of $\delta$ in the experiment may help to improve the fidelity of logic gates. However, due to the constraints imposed by the experimental platform considered, $\Omega_2/2\pi$ cannot exceed 50~MHz, but $\delta$ can be achieved considering the two-photon detuning in the model. Figs.~\ref{fig7} represent the results of the numerical simulation of $r_A$ under an appropriate $\delta$, indicating that choosing an appropriate detuning parameter $\delta$ also contributes to improving the algorithm performance. Compared to the results of the two-photon resonance ($\delta /2\pi=0$~kHz), the performance of FALQON has been improved to some extent in Fig.~\ref{fig7}. Furthermore, we anticipate that, on experimental platforms that allow $\Omega_2/2\pi$ to exceed 50~MHz, the fidelity of controlled-phase gates at small angles will increase, thus further enhancing the performance of FALQON.

\begin{figure}
\includegraphics[width=1\linewidth]{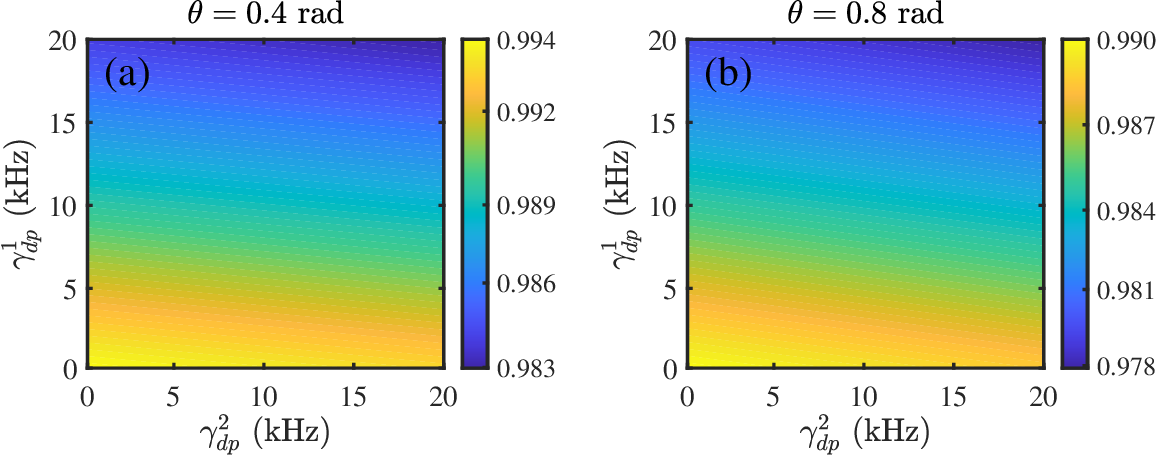}
\caption{The optimally-tuned controlled-phase gates of phase $\theta=0.4$~rad ($\delta /2\pi=-5.6$~kHz) and (b) $\theta=0.8$~rad ($\delta /2\pi=-6.0$~kHz) fidelity with laser phase noise governed by Eqs.~(\ref{eq10}) and ~(\ref{eq18}).}\label{fig8}
\end{figure}

Unlike the fluctuations discussed earlier, the laser phase noise can only have a negative effect on the optimally-tuned controlled-phase gates. Laser phase noise can be expressed as $\Omega_{i}(t)=\Omega_{i}\exp[i\phi_{i}(t)]$, where $\phi_{i}(t)$ is a stochastic process associated with power spectral density $\phi_{i}(t)$ characterized by phase-modulated Fourier frequency $f$. However, directly quantifying laser phase noise is challenging, as $S_{\phi}(f)$ is based on specific experimental test results \cite{de2018analysis,lee2019coherent}. Fortunately, the average effect of laser phase noise leads to Rabi oscillation dephasing \cite{tamura2020analysis,madjarov2020high}, and it can be described in the form of Lindblad operators as
\begin{equation}\label{eq18}
\mathcal{L}_{ln}[\rho]=\sum_{i=a,b}L_{ln}^{(i)}\rho L_{ln}^{(i)\dagger}-\frac{1}{2}\left\{L_{ln}^{(i)\dagger}L_{ln}^{(i)},\rho\right\},
\end{equation}
where $\mathcal{L}_{l1}=\sqrt{\gamma^{1}_{dp}}(|{p}\rangle\langle{p}|-|{1}\rangle\langle{1}|)$ and $\mathcal{L}_{l2}=\sqrt{\gamma^{2}_{dp}}(|{r}\rangle\langle{r}|-|{p}\rangle\langle{p}|)$ describing the dephasing between $|p\rangle$ and $|1\rangle$, and between $|r\rangle$ and $|p\rangle$ caused by the phase noise of $\Omega_{1}(t)$ and $\Omega_2$, respectively. The numerical results in Fig.~\ref{fig8} show the negative effect of two dephasing rates of $\gamma^{1(2)}_{dp}$ $\in$ [0, 20]~kHz on gate fidelity, and the dephasing between $|p\rangle$ and $|1\rangle$ is more influential. We expect that as quantum devices advance, the adverse effects of laser phase noise will diminish, paving the way for high-fidelity small-angle controlled-phase gates.

\section{Conclusion}\label{sec5}
In conclusion, we evaluated an approach for the deployment of a high-fidelity and resilient small-angle controlled-phase gate of Rydberg atoms using an experimental platform based on $^{87}$Rb \cite{fu2022high}. Following that, we investigate its application to combinatorial optimization problems using FALQON. The FALQON algorithm is characterized by its independence from expensive classical optimization resources. However, it requires deeper computational depth to achieve convergence, making it more susceptible to errors, which mainly introduced by two-qubit entanglemet gates. To address this challenge, we propose an improved implementation scheme for FALQON. This scheme not only significantly reduces the required number of two-qubit entanglement gates but also enhances the quality of individual two-qubit entanglemet gate. We investigate the benefits of this method in the case of Max-Cut issues involving two to four qubits. Our results show that our technique improves FALQON's performance, allowing it to outperform the scheme implemented with CZ gates and maintain good convergence performance in different types of gragh even when spontaneous emission is taken into account. We also discuss the resistance of the logic gate to laser intensity fluctuation, detuning fluctuation, and laser phase noise. Therefore, our research holds considerable significance for the implementation and realization of the quantum advantage of FALQON.

While our research primarily focuses on FALQON, the implications of our findings extend to other quantum algorithms reliant on small-angle controlled-phase gates in quantum circuit implementations, such as the quantum Fourier transform and the QAOA. Specifically, when executing a quantum Fourier transform with a larger number $n$ of qubits, it requires controlled phase gates with smaller angles, represented as $\theta=2\pi/2^n$.
Furthermore, the utilization of small-angle controlled-phase gates becomes increasingly probable in the context of the QAOA when tackling more intricate combinatorial optimization problems, as optimal parameters at small angles are more likely to emerge during optimization. Additionally, small-angle controlled-phase gates find application in composite controlled-phase gates, which serve to compensate for errors in the phase of a single gate~\cite{PhysRevA.92.022333}.
Moreover, our model can be extended to various combinatorial optimization issues, including Exact-Cover problems. We anticipate that our findings will significantly contribute to the practical realization of quantum computers and quantum algorithms in near-term neutral-atom systems.

\begin{acknowledgments}
This work was supported by the National Natural Science
Foundation of China (NSFC) under Grant No. 12174048.
JBY acknowledges support from
the National Research Foundation Singapore (NRF2021-
QEP2-02-P01), A*STAR Career Development
Award (C210112010), and A*STAR (C230917003, C230917007).
\end{acknowledgments}

 \bibliography{manuscript.bbl}

\begin{thebibliography}{89}%
\makeatletter
\providecommand \@ifxundefined [1]{%
 \@ifx{#1\undefined}
}%
\providecommand \@ifnum [1]{%
 \ifnum #1\expandafter \@firstoftwo
 \else \expandafter \@secondoftwo
 \fi
}%
\providecommand \@ifx [1]{%
 \ifx #1\expandafter \@firstoftwo
 \else \expandafter \@secondoftwo
 \fi
}%
\providecommand \natexlab [1]{#1}%
\providecommand \enquote  [1]{``#1''}%
\providecommand \bibnamefont  [1]{#1}%
\providecommand \bibfnamefont [1]{#1}%
\providecommand \citenamefont [1]{#1}%
\providecommand \href@noop [0]{\@secondoftwo}%
\providecommand \href [0]{\begingroup \@sanitize@url \@href}%
\providecommand \@href[1]{\@@startlink{#1}\@@href}%
\providecommand \@@href[1]{\endgroup#1\@@endlink}%
\providecommand \@sanitize@url [0]{\catcode `\\12\catcode `\$12\catcode `\&12\catcode `\#12\catcode `\^12\catcode `\_12\catcode `\%12\relax}%
\providecommand \@@startlink[1]{}%
\providecommand \@@endlink[0]{}%
\providecommand \url  [0]{\begingroup\@sanitize@url \@url }%
\providecommand \@url [1]{\endgroup\@href {#1}{\urlprefix }}%
\providecommand \urlprefix  [0]{URL }%
\providecommand \Eprint [0]{\href }%
\providecommand \doibase [0]{http://dx.doi.org/}%
\providecommand \selectlanguage [0]{\@gobble}%
\providecommand \bibinfo  [0]{\@secondoftwo}%
\providecommand \bibfield  [0]{\@secondoftwo}%
\providecommand \translation [1]{[#1]}%
\providecommand \BibitemOpen [0]{}%
\providecommand \bibitemStop [0]{}%
\providecommand \bibitemNoStop [0]{.\EOS\space}%
\providecommand \EOS [0]{\spacefactor3000\relax}%
\providecommand \BibitemShut  [1]{\csname bibitem#1\endcsname}%
\let\auto@bib@innerbib\@empty
\bibitem [{\citenamefont {Korte}\ \emph {et~al.}(2011)\citenamefont {Korte}, \citenamefont {Vygen}, \citenamefont {Korte},\ and\ \citenamefont {Vygen}}]{korte2011combinatorial}%
  \BibitemOpen
  \bibfield  {author} {\bibinfo {author} {\bibfnamefont {Bernhard~H}\ \bibnamefont {Korte}}, \bibinfo {author} {\bibfnamefont {Jens}\ \bibnamefont {Vygen}}, \bibinfo {author} {\bibfnamefont {B}~\bibnamefont {Korte}}, \ and\ \bibinfo {author} {\bibfnamefont {J}~\bibnamefont {Vygen}},\ }\href@noop {} {\emph {\bibinfo {title} {Combinatorial optimization}}},\ Vol.~\bibinfo {volume} {1}\ (\bibinfo  {publisher} {Springer},\ \bibinfo {year} {2011})\BibitemShut {NoStop}%
\bibitem [{\citenamefont {Cao}\ \emph {et~al.}(2019)\citenamefont {Cao}, \citenamefont {Romero}, \citenamefont {Olson}, \citenamefont {Degroote}, \citenamefont {Johnson}, \citenamefont {Kieferov{\'a}}, \citenamefont {Kivlichan}, \citenamefont {Menke}, \citenamefont {Peropadre}, \citenamefont {Sawaya} \emph {et~al.}}]{cao2019quantum}%
  \BibitemOpen
  \bibfield  {author} {\bibinfo {author} {\bibfnamefont {Yudong}\ \bibnamefont {Cao}}, \bibinfo {author} {\bibfnamefont {Jonathan}\ \bibnamefont {Romero}}, \bibinfo {author} {\bibfnamefont {Jonathan~P}\ \bibnamefont {Olson}}, \bibinfo {author} {\bibfnamefont {Matthias}\ \bibnamefont {Degroote}}, \bibinfo {author} {\bibfnamefont {Peter~D}\ \bibnamefont {Johnson}}, \bibinfo {author} {\bibfnamefont {M{\'a}ria}\ \bibnamefont {Kieferov{\'a}}}, \bibinfo {author} {\bibfnamefont {Ian~D}\ \bibnamefont {Kivlichan}}, \bibinfo {author} {\bibfnamefont {Tim}\ \bibnamefont {Menke}}, \bibinfo {author} {\bibfnamefont {Borja}\ \bibnamefont {Peropadre}}, \bibinfo {author} {\bibfnamefont {Nicolas~PD}\ \bibnamefont {Sawaya}},  \emph {et~al.},\ }\bibfield  {title} {\enquote {\bibinfo {title} {Quantum chemistry in the age of quantum computing},}\ }\href {https://doi.org/10.1021/acs.chemrev.8b00803} {\bibfield  {journal} {\bibinfo  {journal} {Chemical reviews}\ }\textbf {\bibinfo {volume} {119}},\ \bibinfo {pages} {10856--10915}
  (\bibinfo {year} {2019})}\BibitemShut {NoStop}%
\bibitem [{\citenamefont {Steiglitz}\ and\ \citenamefont {Papadimitriou}(1998)}]{drug}%
  \BibitemOpen
  \bibfield  {author} {\bibinfo {author} {\bibfnamefont {Kenneth}\ \bibnamefont {Steiglitz}}\ and\ \bibinfo {author} {\bibfnamefont {Christos~H.}\ \bibnamefont {Papadimitriou}},\ }\href {https://cir.nii.ac.jp/crid/1130000797723467648} {\emph {\bibinfo {title} {Combinatorial optimization : algorithms and complexity}}},\ \bibinfo {edition} {dover}\ ed.\ (\bibinfo  {publisher} {Dover Publications},\ \bibinfo {year} {1998})\BibitemShut {NoStop}%
\bibitem [{\citenamefont {Lucas}(2014)}]{Ising}%
  \BibitemOpen
  \bibfield  {author} {\bibinfo {author} {\bibfnamefont {Andrew}\ \bibnamefont {Lucas}},\ }\bibfield  {title} {\enquote {\bibinfo {title} {Ising formulations of many np problems},}\ }\href {\doibase 10.3389/fphy.2014.00005} {\bibfield  {journal} {\bibinfo  {journal} {Frontiers in physics}\ }\textbf {\bibinfo {volume} {2}},\ \bibinfo {pages} {5} (\bibinfo {year} {2014})}\BibitemShut {NoStop}%
\bibitem [{\citenamefont {Magann}\ \emph {et~al.}(2022{\natexlab{a}})\citenamefont {Magann}, \citenamefont {Rudinger}, \citenamefont {Grace},\ and\ \citenamefont {Sarovar}}]{FALQON}%
  \BibitemOpen
  \bibfield  {author} {\bibinfo {author} {\bibfnamefont {Alicia~B.}\ \bibnamefont {Magann}}, \bibinfo {author} {\bibfnamefont {Kenneth~M.}\ \bibnamefont {Rudinger}}, \bibinfo {author} {\bibfnamefont {Matthew~D.}\ \bibnamefont {Grace}}, \ and\ \bibinfo {author} {\bibfnamefont {Mohan}\ \bibnamefont {Sarovar}},\ }\bibfield  {title} {\enquote {\bibinfo {title} {Feedback-based quantum optimization},}\ }\href {\doibase 10.1103/PhysRevLett.129.250502} {\bibfield  {journal} {\bibinfo  {journal} {Phys. Rev. Lett.}\ }\textbf {\bibinfo {volume} {129}},\ \bibinfo {pages} {250502} (\bibinfo {year} {2022}{\natexlab{a}})}\BibitemShut {NoStop}%
\bibitem [{\citenamefont {Magann}\ \emph {et~al.}(2022{\natexlab{b}})\citenamefont {Magann}, \citenamefont {Rudinger}, \citenamefont {Grace},\ and\ \citenamefont {Sarovar}}]{magann2022lyapunov}%
  \BibitemOpen
  \bibfield  {author} {\bibinfo {author} {\bibfnamefont {Alicia~B.}\ \bibnamefont {Magann}}, \bibinfo {author} {\bibfnamefont {Kenneth~M.}\ \bibnamefont {Rudinger}}, \bibinfo {author} {\bibfnamefont {Matthew~D.}\ \bibnamefont {Grace}}, \ and\ \bibinfo {author} {\bibfnamefont {Mohan}\ \bibnamefont {Sarovar}},\ }\bibfield  {title} {\enquote {\bibinfo {title} {Lyapunov-control-inspired strategies for quantum combinatorial optimization},}\ }\href {\doibase 10.1103/PhysRevA.106.062414} {\bibfield  {journal} {\bibinfo  {journal} {Phys. Rev. A}\ }\textbf {\bibinfo {volume} {106}},\ \bibinfo {pages} {062414} (\bibinfo {year} {2022}{\natexlab{b}})}\BibitemShut {NoStop}%
\bibitem [{\citenamefont {Farhi}\ \emph {et~al.}(2014)\citenamefont {Farhi}, \citenamefont {Goldstone},\ and\ \citenamefont {Gutmann}}]{QAOA}%
  \BibitemOpen
  \bibfield  {author} {\bibinfo {author} {\bibfnamefont {Edward}\ \bibnamefont {Farhi}}, \bibinfo {author} {\bibfnamefont {Jeffrey}\ \bibnamefont {Goldstone}}, \ and\ \bibinfo {author} {\bibfnamefont {Sam}\ \bibnamefont {Gutmann}},\ }\bibfield  {title} {\enquote {\bibinfo {title} {A quantum approximate optimization algorithm},}\ }\href {https://arxiv.org/abs/1411.4028} {\bibfield  {journal} {\bibinfo  {journal} {arXiv preprint arXiv:1411.4028}\ } (\bibinfo {year} {2014})}\BibitemShut {NoStop}%
\bibitem [{\citenamefont {Chalupnik}\ \emph {et~al.}(2022)\citenamefont {Chalupnik}, \citenamefont {Melo}, \citenamefont {Alexeev},\ and\ \citenamefont {Galda}}]{9951267}%
  \BibitemOpen
  \bibfield  {author} {\bibinfo {author} {\bibfnamefont {Michelle}\ \bibnamefont {Chalupnik}}, \bibinfo {author} {\bibfnamefont {Hans}\ \bibnamefont {Melo}}, \bibinfo {author} {\bibfnamefont {Yuri}\ \bibnamefont {Alexeev}}, \ and\ \bibinfo {author} {\bibfnamefont {Alexey}\ \bibnamefont {Galda}},\ }\bibfield  {title} {\enquote {\bibinfo {title} {Augmenting qaoa ansatz with multiparameter problem-independent layer},}\ }in\ \href {\doibase 10.1109/QCE53715.2022.00028} {\emph {\bibinfo {booktitle} {2022 IEEE International Conference on Quantum Computing and Engineering (QCE)}}}\ (\bibinfo {year} {2022})\ pp.\ \bibinfo {pages} {97--103}\BibitemShut {NoStop}%
\bibitem [{\citenamefont {Chandarana}\ \emph {et~al.}(2022)\citenamefont {Chandarana}, \citenamefont {Hegade}, \citenamefont {Paul}, \citenamefont {Albarr\'an-Arriagada}, \citenamefont {Solano}, \citenamefont {del Campo},\ and\ \citenamefont {Chen}}]{PhysRevResearch.4.013141}%
  \BibitemOpen
  \bibfield  {author} {\bibinfo {author} {\bibfnamefont {P.}~\bibnamefont {Chandarana}}, \bibinfo {author} {\bibfnamefont {N.~N.}\ \bibnamefont {Hegade}}, \bibinfo {author} {\bibfnamefont {K.}~\bibnamefont {Paul}}, \bibinfo {author} {\bibfnamefont {F.}~\bibnamefont {Albarr\'an-Arriagada}}, \bibinfo {author} {\bibfnamefont {E.}~\bibnamefont {Solano}}, \bibinfo {author} {\bibfnamefont {A.}~\bibnamefont {del Campo}}, \ and\ \bibinfo {author} {\bibfnamefont {Xi}~\bibnamefont {Chen}},\ }\bibfield  {title} {\enquote {\bibinfo {title} {Digitized-counterdiabatic quantum approximate optimization algorithm},}\ }\href {\doibase 10.1103/PhysRevResearch.4.013141} {\bibfield  {journal} {\bibinfo  {journal} {Phys. Rev. Res.}\ }\textbf {\bibinfo {volume} {4}},\ \bibinfo {pages} {013141} (\bibinfo {year} {2022})}\BibitemShut {NoStop}%
\bibitem [{\citenamefont {Yu}\ \emph {et~al.}(2022)\citenamefont {Yu}, \citenamefont {Cao}, \citenamefont {Dewey}, \citenamefont {Wang}, \citenamefont {Shannon},\ and\ \citenamefont {Joynt}}]{PhysRevResearch.4.023249}%
  \BibitemOpen
  \bibfield  {author} {\bibinfo {author} {\bibfnamefont {Yunlong}\ \bibnamefont {Yu}}, \bibinfo {author} {\bibfnamefont {Chenfeng}\ \bibnamefont {Cao}}, \bibinfo {author} {\bibfnamefont {Carter}\ \bibnamefont {Dewey}}, \bibinfo {author} {\bibfnamefont {Xiang-Bin}\ \bibnamefont {Wang}}, \bibinfo {author} {\bibfnamefont {Nic}\ \bibnamefont {Shannon}}, \ and\ \bibinfo {author} {\bibfnamefont {Robert}\ \bibnamefont {Joynt}},\ }\bibfield  {title} {\enquote {\bibinfo {title} {Quantum approximate optimization algorithm with adaptive bias fields},}\ }\href {\doibase 10.1103/PhysRevResearch.4.023249} {\bibfield  {journal} {\bibinfo  {journal} {Phys. Rev. Res.}\ }\textbf {\bibinfo {volume} {4}},\ \bibinfo {pages} {023249} (\bibinfo {year} {2022})}\BibitemShut {NoStop}%
\bibitem [{\citenamefont {Zhu}\ \emph {et~al.}(2022)\citenamefont {Zhu}, \citenamefont {Tang}, \citenamefont {Barron}, \citenamefont {Calderon-Vargas}, \citenamefont {Mayhall}, \citenamefont {Barnes},\ and\ \citenamefont {Economou}}]{PhysRevResearch.4.033029}%
  \BibitemOpen
  \bibfield  {author} {\bibinfo {author} {\bibfnamefont {Linghua}\ \bibnamefont {Zhu}}, \bibinfo {author} {\bibfnamefont {Ho~Lun}\ \bibnamefont {Tang}}, \bibinfo {author} {\bibfnamefont {George~S.}\ \bibnamefont {Barron}}, \bibinfo {author} {\bibfnamefont {F.~A.}\ \bibnamefont {Calderon-Vargas}}, \bibinfo {author} {\bibfnamefont {Nicholas~J.}\ \bibnamefont {Mayhall}}, \bibinfo {author} {\bibfnamefont {Edwin}\ \bibnamefont {Barnes}}, \ and\ \bibinfo {author} {\bibfnamefont {Sophia~E.}\ \bibnamefont {Economou}},\ }\bibfield  {title} {\enquote {\bibinfo {title} {Adaptive quantum approximate optimization algorithm for solving combinatorial problems on a quantum computer},}\ }\href {\doibase 10.1103/PhysRevResearch.4.033029} {\bibfield  {journal} {\bibinfo  {journal} {Phys. Rev. Res.}\ }\textbf {\bibinfo {volume} {4}},\ \bibinfo {pages} {033029} (\bibinfo {year} {2022})}\BibitemShut {NoStop}%
\bibitem [{\citenamefont {Bravyi}\ \emph {et~al.}(2020)\citenamefont {Bravyi}, \citenamefont {Kliesch}, \citenamefont {Koenig},\ and\ \citenamefont {Tang}}]{PhysRevLett.125.260505}%
  \BibitemOpen
  \bibfield  {author} {\bibinfo {author} {\bibfnamefont {Sergey}\ \bibnamefont {Bravyi}}, \bibinfo {author} {\bibfnamefont {Alexander}\ \bibnamefont {Kliesch}}, \bibinfo {author} {\bibfnamefont {Robert}\ \bibnamefont {Koenig}}, \ and\ \bibinfo {author} {\bibfnamefont {Eugene}\ \bibnamefont {Tang}},\ }\bibfield  {title} {\enquote {\bibinfo {title} {Obstacles to variational quantum optimization from symmetry protection},}\ }\href {\doibase 10.1103/PhysRevLett.125.260505} {\bibfield  {journal} {\bibinfo  {journal} {Phys. Rev. Lett.}\ }\textbf {\bibinfo {volume} {125}},\ \bibinfo {pages} {260505} (\bibinfo {year} {2020})}\BibitemShut {NoStop}%
\bibitem [{\citenamefont {Hadfield}\ \emph {et~al.}(2019)\citenamefont {Hadfield}, \citenamefont {Wang}, \citenamefont {O’Gorman}, \citenamefont {Rieffel}, \citenamefont {Venturelli},\ and\ \citenamefont {Biswas}}]{a12020034}%
  \BibitemOpen
  \bibfield  {author} {\bibinfo {author} {\bibfnamefont {Stuart}\ \bibnamefont {Hadfield}}, \bibinfo {author} {\bibfnamefont {Zhihui}\ \bibnamefont {Wang}}, \bibinfo {author} {\bibfnamefont {Bryan}\ \bibnamefont {O’Gorman}}, \bibinfo {author} {\bibfnamefont {Eleanor~G.}\ \bibnamefont {Rieffel}}, \bibinfo {author} {\bibfnamefont {Davide}\ \bibnamefont {Venturelli}}, \ and\ \bibinfo {author} {\bibfnamefont {Rupak}\ \bibnamefont {Biswas}},\ }\bibfield  {title} {\enquote {\bibinfo {title} {From the quantum approximate optimization algorithm to a quantum alternating operator ansatz},}\ }\href {\doibase 10.3390/a12020034} {\bibfield  {journal} {\bibinfo  {journal} {Algorithms}\ }\textbf {\bibinfo {volume} {12}} (\bibinfo {year} {2019}),\ 10.3390/a12020034}\BibitemShut {NoStop}%
\bibitem [{\citenamefont {Egger}\ \emph {et~al.}(2021)\citenamefont {Egger}, \citenamefont {Mare{\v{c}}ek},\ and\ \citenamefont {Woerner}}]{Egger2021warmstartingquantum}%
  \BibitemOpen
  \bibfield  {author} {\bibinfo {author} {\bibfnamefont {Daniel~J.}\ \bibnamefont {Egger}}, \bibinfo {author} {\bibfnamefont {Jakub}\ \bibnamefont {Mare{\v{c}}ek}}, \ and\ \bibinfo {author} {\bibfnamefont {Stefan}\ \bibnamefont {Woerner}},\ }\bibfield  {title} {\enquote {\bibinfo {title} {Warm-starting quantum optimization},}\ }\href {\doibase 10.22331/q-2021-06-17-479} {\bibfield  {journal} {\bibinfo  {journal} {{Quantum}}\ }\textbf {\bibinfo {volume} {5}},\ \bibinfo {pages} {479} (\bibinfo {year} {2021})}\BibitemShut {NoStop}%
\bibitem [{\citenamefont {Yoshioka}\ \emph {et~al.}(2023)\citenamefont {Yoshioka}, \citenamefont {Sasada}, \citenamefont {Nakano},\ and\ \citenamefont {Fujii}}]{PhysRevResearch.5.023071}%
  \BibitemOpen
  \bibfield  {author} {\bibinfo {author} {\bibfnamefont {Takuya}\ \bibnamefont {Yoshioka}}, \bibinfo {author} {\bibfnamefont {Keita}\ \bibnamefont {Sasada}}, \bibinfo {author} {\bibfnamefont {Yuichiro}\ \bibnamefont {Nakano}}, \ and\ \bibinfo {author} {\bibfnamefont {Keisuke}\ \bibnamefont {Fujii}},\ }\bibfield  {title} {\enquote {\bibinfo {title} {Fermionic quantum approximate optimization algorithm},}\ }\href {\doibase 10.1103/PhysRevResearch.5.023071} {\bibfield  {journal} {\bibinfo  {journal} {Phys. Rev. Res.}\ }\textbf {\bibinfo {volume} {5}},\ \bibinfo {pages} {023071} (\bibinfo {year} {2023})}\BibitemShut {NoStop}%
\bibitem [{\citenamefont {Tilly}\ \emph {et~al.}(2022)\citenamefont {Tilly}, \citenamefont {Chen}, \citenamefont {Cao}, \citenamefont {Picozzi}, \citenamefont {Setia}, \citenamefont {Li}, \citenamefont {Grant}, \citenamefont {Wossnig}, \citenamefont {Rungger}, \citenamefont {Booth},\ and\ \citenamefont {Tennyson}}]{TILLY20221}%
  \BibitemOpen
  \bibfield  {author} {\bibinfo {author} {\bibfnamefont {Jules}\ \bibnamefont {Tilly}}, \bibinfo {author} {\bibfnamefont {Hongxiang}\ \bibnamefont {Chen}}, \bibinfo {author} {\bibfnamefont {Shuxiang}\ \bibnamefont {Cao}}, \bibinfo {author} {\bibfnamefont {Dario}\ \bibnamefont {Picozzi}}, \bibinfo {author} {\bibfnamefont {Kanav}\ \bibnamefont {Setia}}, \bibinfo {author} {\bibfnamefont {Ying}\ \bibnamefont {Li}}, \bibinfo {author} {\bibfnamefont {Edward}\ \bibnamefont {Grant}}, \bibinfo {author} {\bibfnamefont {Leonard}\ \bibnamefont {Wossnig}}, \bibinfo {author} {\bibfnamefont {Ivan}\ \bibnamefont {Rungger}}, \bibinfo {author} {\bibfnamefont {George~H.}\ \bibnamefont {Booth}}, \ and\ \bibinfo {author} {\bibfnamefont {Jonathan}\ \bibnamefont {Tennyson}},\ }\bibfield  {title} {\enquote {\bibinfo {title} {The variational quantum eigensolver: A review of methods and best practices},}\ }\href {\doibase https://doi.org/10.1016/j.physrep.2022.08.003} {\bibfield  {journal} {\bibinfo  {journal} {Physics Reports}\
  }\textbf {\bibinfo {volume} {986}},\ \bibinfo {pages} {1--128} (\bibinfo {year} {2022})}\BibitemShut {NoStop}%
\bibitem [{\citenamefont {Holmes}\ \emph {et~al.}(2022)\citenamefont {Holmes}, \citenamefont {Sharma}, \citenamefont {Cerezo},\ and\ \citenamefont {Coles}}]{PRXQuantum.3.010313}%
  \BibitemOpen
  \bibfield  {author} {\bibinfo {author} {\bibfnamefont {Zo\"e}\ \bibnamefont {Holmes}}, \bibinfo {author} {\bibfnamefont {Kunal}\ \bibnamefont {Sharma}}, \bibinfo {author} {\bibfnamefont {M.}~\bibnamefont {Cerezo}}, \ and\ \bibinfo {author} {\bibfnamefont {Patrick~J.}\ \bibnamefont {Coles}},\ }\bibfield  {title} {\enquote {\bibinfo {title} {Connecting ansatz expressibility to gradient magnitudes and barren plateaus},}\ }\href {\doibase 10.1103/PRXQuantum.3.010313} {\bibfield  {journal} {\bibinfo  {journal} {PRX Quantum}\ }\textbf {\bibinfo {volume} {3}},\ \bibinfo {pages} {010313} (\bibinfo {year} {2022})}\BibitemShut {NoStop}%
\bibitem [{\citenamefont {Guerreschi}\ and\ \citenamefont {Matsuura}(2019)}]{guerreschi2019qaoa}%
  \BibitemOpen
  \bibfield  {author} {\bibinfo {author} {\bibfnamefont {Gian~Giacomo}\ \bibnamefont {Guerreschi}}\ and\ \bibinfo {author} {\bibfnamefont {Anne~Y}\ \bibnamefont {Matsuura}},\ }\bibfield  {title} {\enquote {\bibinfo {title} {Qaoa for max-cut requires hundreds of qubits for quantum speed-up},}\ }\href {https://doi.org/10.1038/s41598-019-43176-9} {\bibfield  {journal} {\bibinfo  {journal} {Scientific reports}\ }\textbf {\bibinfo {volume} {9}},\ \bibinfo {pages} {6903} (\bibinfo {year} {2019})}\BibitemShut {NoStop}%
\bibitem [{\citenamefont {Bärtschi}\ and\ \citenamefont {Eidenbenz}(2020)}]{9259965}%
  \BibitemOpen
  \bibfield  {author} {\bibinfo {author} {\bibfnamefont {Andreas}\ \bibnamefont {Bärtschi}}\ and\ \bibinfo {author} {\bibfnamefont {Stephan}\ \bibnamefont {Eidenbenz}},\ }\bibfield  {title} {\enquote {\bibinfo {title} {Grover mixers for qaoa: Shifting complexity from mixer design to state preparation},}\ }in\ \href {\doibase 10.1109/QCE49297.2020.00020} {\emph {\bibinfo {booktitle} {2020 IEEE International Conference on Quantum Computing and Engineering (QCE)}}}\ (\bibinfo {year} {2020})\ pp.\ \bibinfo {pages} {72--82}\BibitemShut {NoStop}%
\bibitem [{\citenamefont {Fuchs}\ \emph {et~al.}(2021)\citenamefont {Fuchs}, \citenamefont {Kolden}, \citenamefont {Aase},\ and\ \citenamefont {Sartor}}]{fuchs2021efficient}%
  \BibitemOpen
  \bibfield  {author} {\bibinfo {author} {\bibfnamefont {Franz~G}\ \bibnamefont {Fuchs}}, \bibinfo {author} {\bibfnamefont {Herman~{\O}ie}\ \bibnamefont {Kolden}}, \bibinfo {author} {\bibfnamefont {Niels~Henrik}\ \bibnamefont {Aase}}, \ and\ \bibinfo {author} {\bibfnamefont {Giorgio}\ \bibnamefont {Sartor}},\ }\bibfield  {title} {\enquote {\bibinfo {title} {Efficient encoding of the weighted max k-cut on a quantum computer using qaoa},}\ }\href {https://link.springer.com/article/10.1007/s42979-020-00437-z} {\bibfield  {journal} {\bibinfo  {journal} {SN Computer Science}\ }\textbf {\bibinfo {volume} {2}},\ \bibinfo {pages} {89} (\bibinfo {year} {2021})}\BibitemShut {NoStop}%
\bibitem [{\citenamefont {Chancellor}(2019)}]{chancellor2019domain}%
  \BibitemOpen
  \bibfield  {author} {\bibinfo {author} {\bibfnamefont {Nicholas}\ \bibnamefont {Chancellor}},\ }\bibfield  {title} {\enquote {\bibinfo {title} {Domain wall encoding of discrete variables for quantum annealing and qaoa},}\ }\href {https://iopscience.iop.org/article/10.1088/2058-9565/ab33c2/meta} {\bibfield  {journal} {\bibinfo  {journal} {Quantum Science and Technology}\ }\textbf {\bibinfo {volume} {4}},\ \bibinfo {pages} {045004} (\bibinfo {year} {2019})}\BibitemShut {NoStop}%
\bibitem [{\citenamefont {Cerezo}\ \emph {et~al.}(2021{\natexlab{a}})\citenamefont {Cerezo}, \citenamefont {Arrasmith}, \citenamefont {Babbush}, \citenamefont {Benjamin}, \citenamefont {Endo}, \citenamefont {Fujii}, \citenamefont {McClean}, \citenamefont {Mitarai}, \citenamefont {Yuan}, \citenamefont {Cincio} \emph {et~al.}}]{cerezo2021variational}%
  \BibitemOpen
  \bibfield  {author} {\bibinfo {author} {\bibfnamefont {Marco}\ \bibnamefont {Cerezo}}, \bibinfo {author} {\bibfnamefont {Andrew}\ \bibnamefont {Arrasmith}}, \bibinfo {author} {\bibfnamefont {Ryan}\ \bibnamefont {Babbush}}, \bibinfo {author} {\bibfnamefont {Simon~C}\ \bibnamefont {Benjamin}}, \bibinfo {author} {\bibfnamefont {Suguru}\ \bibnamefont {Endo}}, \bibinfo {author} {\bibfnamefont {Keisuke}\ \bibnamefont {Fujii}}, \bibinfo {author} {\bibfnamefont {Jarrod~R}\ \bibnamefont {McClean}}, \bibinfo {author} {\bibfnamefont {Kosuke}\ \bibnamefont {Mitarai}}, \bibinfo {author} {\bibfnamefont {Xiao}\ \bibnamefont {Yuan}}, \bibinfo {author} {\bibfnamefont {Lukasz}\ \bibnamefont {Cincio}},  \emph {et~al.},\ }\bibfield  {title} {\enquote {\bibinfo {title} {Variational quantum algorithms},}\ }\href {https://www.nature.com/articles/s42254-021-00348-9} {\bibfield  {journal} {\bibinfo  {journal} {Nature Reviews Physics}\ }\textbf {\bibinfo {volume} {3}},\ \bibinfo {pages} {625--644} (\bibinfo {year}
  {2021}{\natexlab{a}})}\BibitemShut {NoStop}%
\bibitem [{\citenamefont {Harrigan}\ \emph {et~al.}(2021)\citenamefont {Harrigan}, \citenamefont {Sung}, \citenamefont {Neeley}, \citenamefont {Satzinger}, \citenamefont {Arute}, \citenamefont {Arya}, \citenamefont {Atalaya}, \citenamefont {Bardin}, \citenamefont {Barends}, \citenamefont {Boixo} \emph {et~al.}}]{harrigan2021quantum}%
  \BibitemOpen
  \bibfield  {author} {\bibinfo {author} {\bibfnamefont {Matthew~P}\ \bibnamefont {Harrigan}}, \bibinfo {author} {\bibfnamefont {Kevin~J}\ \bibnamefont {Sung}}, \bibinfo {author} {\bibfnamefont {Matthew}\ \bibnamefont {Neeley}}, \bibinfo {author} {\bibfnamefont {Kevin~J}\ \bibnamefont {Satzinger}}, \bibinfo {author} {\bibfnamefont {Frank}\ \bibnamefont {Arute}}, \bibinfo {author} {\bibfnamefont {Kunal}\ \bibnamefont {Arya}}, \bibinfo {author} {\bibfnamefont {Juan}\ \bibnamefont {Atalaya}}, \bibinfo {author} {\bibfnamefont {Joseph~C}\ \bibnamefont {Bardin}}, \bibinfo {author} {\bibfnamefont {Rami}\ \bibnamefont {Barends}}, \bibinfo {author} {\bibfnamefont {Sergio}\ \bibnamefont {Boixo}},  \emph {et~al.},\ }\bibfield  {title} {\enquote {\bibinfo {title} {Quantum approximate optimization of non-planar graph problems on a planar superconducting processor},}\ }\href {https://www.nature.com/articles/s41567-020-01105-y} {\bibfield  {journal} {\bibinfo  {journal} {Nature Physics}\ }\textbf {\bibinfo {volume} {17}},\
  \bibinfo {pages} {332--336} (\bibinfo {year} {2021})}\BibitemShut {NoStop}%
\bibitem [{\citenamefont {Akshay}\ \emph {et~al.}(2020)\citenamefont {Akshay}, \citenamefont {Philathong}, \citenamefont {Morales},\ and\ \citenamefont {Biamonte}}]{PhysRevLett.124.090504}%
  \BibitemOpen
  \bibfield  {author} {\bibinfo {author} {\bibfnamefont {V.}~\bibnamefont {Akshay}}, \bibinfo {author} {\bibfnamefont {H.}~\bibnamefont {Philathong}}, \bibinfo {author} {\bibfnamefont {M.~E.~S.}\ \bibnamefont {Morales}}, \ and\ \bibinfo {author} {\bibfnamefont {J.~D.}\ \bibnamefont {Biamonte}},\ }\bibfield  {title} {\enquote {\bibinfo {title} {Reachability deficits in quantum approximate optimization},}\ }\href {\doibase 10.1103/PhysRevLett.124.090504} {\bibfield  {journal} {\bibinfo  {journal} {Phys. Rev. Lett.}\ }\textbf {\bibinfo {volume} {124}},\ \bibinfo {pages} {090504} (\bibinfo {year} {2020})}\BibitemShut {NoStop}%
\bibitem [{\citenamefont {Farhi}\ \emph {et~al.}(2022)\citenamefont {Farhi}, \citenamefont {Goldstone}, \citenamefont {Gutmann},\ and\ \citenamefont {Zhou}}]{farhi2022quantum}%
  \BibitemOpen
  \bibfield  {author} {\bibinfo {author} {\bibfnamefont {Edward}\ \bibnamefont {Farhi}}, \bibinfo {author} {\bibfnamefont {Jeffrey}\ \bibnamefont {Goldstone}}, \bibinfo {author} {\bibfnamefont {Sam}\ \bibnamefont {Gutmann}}, \ and\ \bibinfo {author} {\bibfnamefont {Leo}\ \bibnamefont {Zhou}},\ }\bibfield  {title} {\enquote {\bibinfo {title} {The quantum approximate optimization algorithm and the sherrington-kirkpatrick model at infinite size},}\ }\href {https://quantum-journal.org/papers/q-2022-07-07-759/} {\bibfield  {journal} {\bibinfo  {journal} {Quantum}\ }\textbf {\bibinfo {volume} {6}},\ \bibinfo {pages} {759} (\bibinfo {year} {2022})}\BibitemShut {NoStop}%
\bibitem [{\citenamefont {Lloyd}(2018)}]{lloyd2018quantum}%
  \BibitemOpen
  \bibfield  {author} {\bibinfo {author} {\bibfnamefont {Seth}\ \bibnamefont {Lloyd}},\ }\bibfield  {title} {\enquote {\bibinfo {title} {Quantum approximate optimization is computationally universal},}\ }\href {https://arxiv.org/abs/1812.11075} {\bibfield  {journal} {\bibinfo  {journal} {arXiv preprint arXiv:1812.11075}\ } (\bibinfo {year} {2018})}\BibitemShut {NoStop}%
\bibitem [{\citenamefont {Farhi}\ and\ \citenamefont {Harrow}(2016)}]{farhi2016quantum}%
  \BibitemOpen
  \bibfield  {author} {\bibinfo {author} {\bibfnamefont {Edward}\ \bibnamefont {Farhi}}\ and\ \bibinfo {author} {\bibfnamefont {Aram~W}\ \bibnamefont {Harrow}},\ }\bibfield  {title} {\enquote {\bibinfo {title} {Quantum supremacy through the quantum approximate optimization algorithm},}\ }\href {https://arxiv.org/abs/1602.07674} {\bibfield  {journal} {\bibinfo  {journal} {arXiv preprint arXiv:1602.07674}\ } (\bibinfo {year} {2016})}\BibitemShut {NoStop}%
\bibitem [{\citenamefont {Pagano}\ \emph {et~al.}(2020)\citenamefont {Pagano}, \citenamefont {Bapat}, \citenamefont {Becker}, \citenamefont {Collins}, \citenamefont {De}, \citenamefont {Hess}, \citenamefont {Kaplan}, \citenamefont {Kyprianidis}, \citenamefont {Tan}, \citenamefont {Baldwin} \emph {et~al.}}]{pagano2020quantum}%
  \BibitemOpen
  \bibfield  {author} {\bibinfo {author} {\bibfnamefont {Guido}\ \bibnamefont {Pagano}}, \bibinfo {author} {\bibfnamefont {Aniruddha}\ \bibnamefont {Bapat}}, \bibinfo {author} {\bibfnamefont {Patrick}\ \bibnamefont {Becker}}, \bibinfo {author} {\bibfnamefont {Katherine~S}\ \bibnamefont {Collins}}, \bibinfo {author} {\bibfnamefont {Arinjoy}\ \bibnamefont {De}}, \bibinfo {author} {\bibfnamefont {Paul~W}\ \bibnamefont {Hess}}, \bibinfo {author} {\bibfnamefont {Harvey~B}\ \bibnamefont {Kaplan}}, \bibinfo {author} {\bibfnamefont {Antonis}\ \bibnamefont {Kyprianidis}}, \bibinfo {author} {\bibfnamefont {Wen~Lin}\ \bibnamefont {Tan}}, \bibinfo {author} {\bibfnamefont {Christopher}\ \bibnamefont {Baldwin}},  \emph {et~al.},\ }\bibfield  {title} {\enquote {\bibinfo {title} {Quantum approximate optimization of the long-range ising model with a trapped-ion quantum simulator},}\ }\href {https://www.pnas.org/doi/abs/10.1073/pnas.2006373117} {\bibfield  {journal} {\bibinfo  {journal} {Proceedings of the National Academy of
  Sciences}\ }\textbf {\bibinfo {volume} {117}},\ \bibinfo {pages} {25396--25401} (\bibinfo {year} {2020})}\BibitemShut {NoStop}%
\bibitem [{\citenamefont {Vikst\aa{}l}\ \emph {et~al.}(2020)\citenamefont {Vikst\aa{}l}, \citenamefont {Gr\"onkvist}, \citenamefont {Svensson}, \citenamefont {Andersson}, \citenamefont {Johansson},\ and\ \citenamefont {Ferrini}}]{PhysRevApplied.14.034009}%
  \BibitemOpen
  \bibfield  {author} {\bibinfo {author} {\bibfnamefont {Pontus}\ \bibnamefont {Vikst\aa{}l}}, \bibinfo {author} {\bibfnamefont {Mattias}\ \bibnamefont {Gr\"onkvist}}, \bibinfo {author} {\bibfnamefont {Marika}\ \bibnamefont {Svensson}}, \bibinfo {author} {\bibfnamefont {Martin}\ \bibnamefont {Andersson}}, \bibinfo {author} {\bibfnamefont {G\"oran}\ \bibnamefont {Johansson}}, \ and\ \bibinfo {author} {\bibfnamefont {Giulia}\ \bibnamefont {Ferrini}},\ }\bibfield  {title} {\enquote {\bibinfo {title} {Applying the quantum approximate optimization algorithm to the tail-assignment problem},}\ }\href {\doibase 10.1103/PhysRevApplied.14.034009} {\bibfield  {journal} {\bibinfo  {journal} {Phys. Rev. Appl.}\ }\textbf {\bibinfo {volume} {14}},\ \bibinfo {pages} {034009} (\bibinfo {year} {2020})}\BibitemShut {NoStop}%
\bibitem [{\citenamefont {Shaydulin}\ \emph {et~al.}(2019)\citenamefont {Shaydulin}, \citenamefont {Safro},\ and\ \citenamefont {Larson}}]{shaydulin2019multistart}%
  \BibitemOpen
  \bibfield  {author} {\bibinfo {author} {\bibfnamefont {Ruslan}\ \bibnamefont {Shaydulin}}, \bibinfo {author} {\bibfnamefont {Ilya}\ \bibnamefont {Safro}}, \ and\ \bibinfo {author} {\bibfnamefont {Jeffrey}\ \bibnamefont {Larson}},\ }\bibfield  {title} {\enquote {\bibinfo {title} {Multistart methods for quantum approximate optimization},}\ }in\ \href {https://ieeexplore.ieee.org/abstract/document/8916288} {\emph {\bibinfo {booktitle} {2019 IEEE high performance extreme computing conference (HPEC)}}}\ (\bibinfo {organization} {IEEE},\ \bibinfo {year} {2019})\ pp.\ \bibinfo {pages} {1--8}\BibitemShut {NoStop}%
\bibitem [{\citenamefont {Graham}\ \emph {et~al.}(2022)\citenamefont {Graham}, \citenamefont {Song}, \citenamefont {Scott}, \citenamefont {Poole}, \citenamefont {Phuttitarn}, \citenamefont {Jooya}, \citenamefont {Eichler}, \citenamefont {Jiang}, \citenamefont {Marra}, \citenamefont {Grinkemeyer} \emph {et~al.}}]{graham2022multi}%
  \BibitemOpen
  \bibfield  {author} {\bibinfo {author} {\bibfnamefont {TM}~\bibnamefont {Graham}}, \bibinfo {author} {\bibfnamefont {Y}~\bibnamefont {Song}}, \bibinfo {author} {\bibfnamefont {J}~\bibnamefont {Scott}}, \bibinfo {author} {\bibfnamefont {C}~\bibnamefont {Poole}}, \bibinfo {author} {\bibfnamefont {L}~\bibnamefont {Phuttitarn}}, \bibinfo {author} {\bibfnamefont {K}~\bibnamefont {Jooya}}, \bibinfo {author} {\bibfnamefont {P}~\bibnamefont {Eichler}}, \bibinfo {author} {\bibfnamefont {X}~\bibnamefont {Jiang}}, \bibinfo {author} {\bibfnamefont {A}~\bibnamefont {Marra}}, \bibinfo {author} {\bibfnamefont {B}~\bibnamefont {Grinkemeyer}},  \emph {et~al.},\ }\bibfield  {title} {\enquote {\bibinfo {title} {Multi-qubit entanglement and algorithms on a neutral-atom quantum computer},}\ }\href {https://www.nature.com/articles/s41586-022-04603-6} {\bibfield  {journal} {\bibinfo  {journal} {Nature}\ }\textbf {\bibinfo {volume} {604}},\ \bibinfo {pages} {457--462} (\bibinfo {year} {2022})}\BibitemShut {NoStop}%
\bibitem [{\citenamefont {Wang}\ \emph {et~al.}(2021)\citenamefont {Wang}, \citenamefont {Fontana}, \citenamefont {Cerezo}, \citenamefont {Sharma}, \citenamefont {Sone}, \citenamefont {Cincio},\ and\ \citenamefont {Coles}}]{wang2021noise}%
  \BibitemOpen
  \bibfield  {author} {\bibinfo {author} {\bibfnamefont {Samson}\ \bibnamefont {Wang}}, \bibinfo {author} {\bibfnamefont {Enrico}\ \bibnamefont {Fontana}}, \bibinfo {author} {\bibfnamefont {Marco}\ \bibnamefont {Cerezo}}, \bibinfo {author} {\bibfnamefont {Kunal}\ \bibnamefont {Sharma}}, \bibinfo {author} {\bibfnamefont {Akira}\ \bibnamefont {Sone}}, \bibinfo {author} {\bibfnamefont {Lukasz}\ \bibnamefont {Cincio}}, \ and\ \bibinfo {author} {\bibfnamefont {Patrick~J}\ \bibnamefont {Coles}},\ }\bibfield  {title} {\enquote {\bibinfo {title} {Noise-induced barren plateaus in variational quantum algorithms},}\ }\href {https://www.nature.com/articles/s41467-021-27045-6} {\bibfield  {journal} {\bibinfo  {journal} {Nature communications}\ }\textbf {\bibinfo {volume} {12}},\ \bibinfo {pages} {6961} (\bibinfo {year} {2021})}\BibitemShut {NoStop}%
\bibitem [{\citenamefont {Cerezo}\ \emph {et~al.}(2021{\natexlab{b}})\citenamefont {Cerezo}, \citenamefont {Sone}, \citenamefont {Volkoff}, \citenamefont {Cincio},\ and\ \citenamefont {Coles}}]{cerezo2021cost}%
  \BibitemOpen
  \bibfield  {author} {\bibinfo {author} {\bibfnamefont {Marco}\ \bibnamefont {Cerezo}}, \bibinfo {author} {\bibfnamefont {Akira}\ \bibnamefont {Sone}}, \bibinfo {author} {\bibfnamefont {Tyler}\ \bibnamefont {Volkoff}}, \bibinfo {author} {\bibfnamefont {Lukasz}\ \bibnamefont {Cincio}}, \ and\ \bibinfo {author} {\bibfnamefont {Patrick~J}\ \bibnamefont {Coles}},\ }\bibfield  {title} {\enquote {\bibinfo {title} {Cost function dependent barren plateaus in shallow parametrized quantum circuits},}\ }\href {https://www.nature.com/articles/s41467-021-21728-w} {\bibfield  {journal} {\bibinfo  {journal} {Nature communications}\ }\textbf {\bibinfo {volume} {12}},\ \bibinfo {pages} {1791} (\bibinfo {year} {2021}{\natexlab{b}})}\BibitemShut {NoStop}%
\bibitem [{\citenamefont {Stilck~Fran{\c{c}}a}\ and\ \citenamefont {Garcia-Patron}(2021)}]{stilck2021limitations}%
  \BibitemOpen
  \bibfield  {author} {\bibinfo {author} {\bibfnamefont {Daniel}\ \bibnamefont {Stilck~Fran{\c{c}}a}}\ and\ \bibinfo {author} {\bibfnamefont {Raul}\ \bibnamefont {Garcia-Patron}},\ }\bibfield  {title} {\enquote {\bibinfo {title} {Limitations of optimization algorithms on noisy quantum devices},}\ }\href {https://www.nature.com/articles/s41567-021-01356-3} {\bibfield  {journal} {\bibinfo  {journal} {Nature Physics}\ }\textbf {\bibinfo {volume} {17}},\ \bibinfo {pages} {1221--1227} (\bibinfo {year} {2021})}\BibitemShut {NoStop}%
\bibitem [{\citenamefont {Higgott}\ \emph {et~al.}(2019)\citenamefont {Higgott}, \citenamefont {Wang},\ and\ \citenamefont {Brierley}}]{higgott2019variational}%
  \BibitemOpen
  \bibfield  {author} {\bibinfo {author} {\bibfnamefont {Oscar}\ \bibnamefont {Higgott}}, \bibinfo {author} {\bibfnamefont {Daochen}\ \bibnamefont {Wang}}, \ and\ \bibinfo {author} {\bibfnamefont {Stephen}\ \bibnamefont {Brierley}},\ }\bibfield  {title} {\enquote {\bibinfo {title} {Variational quantum computation of excited states},}\ }\href {https://quantum-journal.org/papers/q-2019-07-01-156/?ref=https://githubhelp.com} {\bibfield  {journal} {\bibinfo  {journal} {Quantum}\ }\textbf {\bibinfo {volume} {3}},\ \bibinfo {pages} {156} (\bibinfo {year} {2019})}\BibitemShut {NoStop}%
\bibitem [{\citenamefont {Chen}\ \emph {et~al.}(2020)\citenamefont {Chen}, \citenamefont {Yang}, \citenamefont {Qi}, \citenamefont {Chen}, \citenamefont {Ma},\ and\ \citenamefont {Goan}}]{chen2020variational}%
  \BibitemOpen
  \bibfield  {author} {\bibinfo {author} {\bibfnamefont {Samuel Yen-Chi}\ \bibnamefont {Chen}}, \bibinfo {author} {\bibfnamefont {Chao-Han~Huck}\ \bibnamefont {Yang}}, \bibinfo {author} {\bibfnamefont {Jun}\ \bibnamefont {Qi}}, \bibinfo {author} {\bibfnamefont {Pin-Yu}\ \bibnamefont {Chen}}, \bibinfo {author} {\bibfnamefont {Xiaoli}\ \bibnamefont {Ma}}, \ and\ \bibinfo {author} {\bibfnamefont {Hsi-Sheng}\ \bibnamefont {Goan}},\ }\bibfield  {title} {\enquote {\bibinfo {title} {Variational quantum circuits for deep reinforcement learning},}\ }\href {https://ieeexplore.ieee.org/abstract/document/9144562} {\bibfield  {journal} {\bibinfo  {journal} {IEEE Access}\ }\textbf {\bibinfo {volume} {8}},\ \bibinfo {pages} {141007--141024} (\bibinfo {year} {2020})}\BibitemShut {NoStop}%
\bibitem [{\citenamefont {Khatri}\ \emph {et~al.}(2019)\citenamefont {Khatri}, \citenamefont {LaRose}, \citenamefont {Poremba}, \citenamefont {Cincio}, \citenamefont {Sornborger},\ and\ \citenamefont {Coles}}]{khatri2019quantum}%
  \BibitemOpen
  \bibfield  {author} {\bibinfo {author} {\bibfnamefont {Sumeet}\ \bibnamefont {Khatri}}, \bibinfo {author} {\bibfnamefont {Ryan}\ \bibnamefont {LaRose}}, \bibinfo {author} {\bibfnamefont {Alexander}\ \bibnamefont {Poremba}}, \bibinfo {author} {\bibfnamefont {Lukasz}\ \bibnamefont {Cincio}}, \bibinfo {author} {\bibfnamefont {Andrew~T}\ \bibnamefont {Sornborger}}, \ and\ \bibinfo {author} {\bibfnamefont {Patrick~J}\ \bibnamefont {Coles}},\ }\bibfield  {title} {\enquote {\bibinfo {title} {Quantum-assisted quantum compiling},}\ }\href {https://quantum-journal.org/papers/q-2019-05-13-140/} {\bibfield  {journal} {\bibinfo  {journal} {Quantum}\ }\textbf {\bibinfo {volume} {3}},\ \bibinfo {pages} {140} (\bibinfo {year} {2019})}\BibitemShut {NoStop}%
\bibitem [{\citenamefont {Ebadi}\ \emph {et~al.}(2022)\citenamefont {Ebadi}, \citenamefont {Keesling}, \citenamefont {Cain}, \citenamefont {Wang}, \citenamefont {Levine}, \citenamefont {Bluvstein}, \citenamefont {Semeghini}, \citenamefont {Omran}, \citenamefont {Liu}, \citenamefont {Samajdar}, \citenamefont {Luo}, \citenamefont {Nash}, \citenamefont {Gao}, \citenamefont {Barak}, \citenamefont {Farhi}, \citenamefont {Sachdev}, \citenamefont {Gemelke}, \citenamefont {Zhou}, \citenamefont {Choi}, \citenamefont {Pichler}, \citenamefont {Wang}, \citenamefont {Greiner}, \citenamefont {Vuletić},\ and\ \citenamefont {Lukin}}]{doi:10.1126/science.abo6587}%
  \BibitemOpen
  \bibfield  {author} {\bibinfo {author} {\bibfnamefont {S.}~\bibnamefont {Ebadi}}, \bibinfo {author} {\bibfnamefont {A.}~\bibnamefont {Keesling}}, \bibinfo {author} {\bibfnamefont {M.}~\bibnamefont {Cain}}, \bibinfo {author} {\bibfnamefont {T.~T.}\ \bibnamefont {Wang}}, \bibinfo {author} {\bibfnamefont {H.}~\bibnamefont {Levine}}, \bibinfo {author} {\bibfnamefont {D.}~\bibnamefont {Bluvstein}}, \bibinfo {author} {\bibfnamefont {G.}~\bibnamefont {Semeghini}}, \bibinfo {author} {\bibfnamefont {A.}~\bibnamefont {Omran}}, \bibinfo {author} {\bibfnamefont {J.-G.}\ \bibnamefont {Liu}}, \bibinfo {author} {\bibfnamefont {R.}~\bibnamefont {Samajdar}}, \bibinfo {author} {\bibfnamefont {X.-Z.}\ \bibnamefont {Luo}}, \bibinfo {author} {\bibfnamefont {B.}~\bibnamefont {Nash}}, \bibinfo {author} {\bibfnamefont {X.}~\bibnamefont {Gao}}, \bibinfo {author} {\bibfnamefont {B.}~\bibnamefont {Barak}}, \bibinfo {author} {\bibfnamefont {E.}~\bibnamefont {Farhi}}, \bibinfo {author} {\bibfnamefont {S.}~\bibnamefont {Sachdev}},
  \bibinfo {author} {\bibfnamefont {N.}~\bibnamefont {Gemelke}}, \bibinfo {author} {\bibfnamefont {L.}~\bibnamefont {Zhou}}, \bibinfo {author} {\bibfnamefont {S.}~\bibnamefont {Choi}}, \bibinfo {author} {\bibfnamefont {H.}~\bibnamefont {Pichler}}, \bibinfo {author} {\bibfnamefont {S.-T.}\ \bibnamefont {Wang}}, \bibinfo {author} {\bibfnamefont {M.}~\bibnamefont {Greiner}}, \bibinfo {author} {\bibfnamefont {V.}~\bibnamefont {Vuletić}}, \ and\ \bibinfo {author} {\bibfnamefont {M.~D.}\ \bibnamefont {Lukin}},\ }\bibfield  {title} {\enquote {\bibinfo {title} {Quantum optimization of maximum independent set using rydberg atom arrays},}\ }\href {\doibase 10.1126/science.abo6587} {\bibfield  {journal} {\bibinfo  {journal} {Science}\ }\textbf {\bibinfo {volume} {376}},\ \bibinfo {pages} {1209--1215} (\bibinfo {year} {2022})}\BibitemShut {NoStop}%
\bibitem [{\citenamefont {Galindo}\ and\ \citenamefont {Mart\'{\i}n-Delgado}(2002)}]{galindo2002information}%
  \BibitemOpen
  \bibfield  {author} {\bibinfo {author} {\bibfnamefont {A.}~\bibnamefont {Galindo}}\ and\ \bibinfo {author} {\bibfnamefont {M.~A.}\ \bibnamefont {Mart\'{\i}n-Delgado}},\ }\bibfield  {title} {\enquote {\bibinfo {title} {Information and computation: Classical and quantum aspects},}\ }\href {\doibase 10.1103/RevModPhys.74.347} {\bibfield  {journal} {\bibinfo  {journal} {Rev. Mod. Phys.}\ }\textbf {\bibinfo {volume} {74}},\ \bibinfo {pages} {347--423} (\bibinfo {year} {2002})}\BibitemShut {NoStop}%
\bibitem [{\citenamefont {Ladd}\ \emph {et~al.}(2010)\citenamefont {Ladd}, \citenamefont {Jelezko}, \citenamefont {Laflamme}, \citenamefont {Nakamura}, \citenamefont {Monroe},\ and\ \citenamefont {O’Brien}}]{ladd2010quantum}%
  \BibitemOpen
  \bibfield  {author} {\bibinfo {author} {\bibfnamefont {Thaddeus~D}\ \bibnamefont {Ladd}}, \bibinfo {author} {\bibfnamefont {Fedor}\ \bibnamefont {Jelezko}}, \bibinfo {author} {\bibfnamefont {Raymond}\ \bibnamefont {Laflamme}}, \bibinfo {author} {\bibfnamefont {Yasunobu}\ \bibnamefont {Nakamura}}, \bibinfo {author} {\bibfnamefont {Christopher}\ \bibnamefont {Monroe}}, \ and\ \bibinfo {author} {\bibfnamefont {Jeremy~Lloyd}\ \bibnamefont {O’Brien}},\ }\bibfield  {title} {\enquote {\bibinfo {title} {Quantum computers},}\ }\href {https://doi.org/10.1038/nature08812} {\bibfield  {journal} {\bibinfo  {journal} {nature}\ }\textbf {\bibinfo {volume} {464}},\ \bibinfo {pages} {45--53} (\bibinfo {year} {2010})}\BibitemShut {NoStop}%
\bibitem [{\citenamefont {Wendin}(2017)}]{wendin2017quantum}%
  \BibitemOpen
  \bibfield  {author} {\bibinfo {author} {\bibfnamefont {G{\"o}ran}\ \bibnamefont {Wendin}},\ }\bibfield  {title} {\enquote {\bibinfo {title} {Quantum information processing with superconducting circuits: a review},}\ }\href {https://iopscience.iop.org/article/10.1088/1361-6633/aa7e1a} {\bibfield  {journal} {\bibinfo  {journal} {Reports on Progress in Physics}\ }\textbf {\bibinfo {volume} {80}},\ \bibinfo {pages} {106001} (\bibinfo {year} {2017})}\BibitemShut {NoStop}%
\bibitem [{\citenamefont {Calderon-Vargas}\ \emph {et~al.}(2019)\citenamefont {Calderon-Vargas}, \citenamefont {Barron}, \citenamefont {Deng}, \citenamefont {Sigillito}, \citenamefont {Barnes},\ and\ \citenamefont {Economou}}]{calderon2019fast}%
  \BibitemOpen
  \bibfield  {author} {\bibinfo {author} {\bibfnamefont {F.~A.}\ \bibnamefont {Calderon-Vargas}}, \bibinfo {author} {\bibfnamefont {George~S.}\ \bibnamefont {Barron}}, \bibinfo {author} {\bibfnamefont {Xiu-Hao}\ \bibnamefont {Deng}}, \bibinfo {author} {\bibfnamefont {A.~J.}\ \bibnamefont {Sigillito}}, \bibinfo {author} {\bibfnamefont {Edwin}\ \bibnamefont {Barnes}}, \ and\ \bibinfo {author} {\bibfnamefont {Sophia~E.}\ \bibnamefont {Economou}},\ }\bibfield  {title} {\enquote {\bibinfo {title} {Fast high-fidelity entangling gates for spin qubits in si double quantum dots},}\ }\href {\doibase 10.1103/PhysRevB.100.035304} {\bibfield  {journal} {\bibinfo  {journal} {Phys. Rev. B}\ }\textbf {\bibinfo {volume} {100}},\ \bibinfo {pages} {035304} (\bibinfo {year} {2019})}\BibitemShut {NoStop}%
\bibitem [{\citenamefont {Kanaar}\ \emph {et~al.}(2021)\citenamefont {Kanaar}, \citenamefont {Wolin}, \citenamefont {G\"ung\"ord\"u},\ and\ \citenamefont {Kestner}}]{kanaar2021single}%
  \BibitemOpen
  \bibfield  {author} {\bibinfo {author} {\bibfnamefont {David~W.}\ \bibnamefont {Kanaar}}, \bibinfo {author} {\bibfnamefont {Sidney}\ \bibnamefont {Wolin}}, \bibinfo {author} {\bibfnamefont {Utkan}\ \bibnamefont {G\"ung\"ord\"u}}, \ and\ \bibinfo {author} {\bibfnamefont {J.~P.}\ \bibnamefont {Kestner}},\ }\bibfield  {title} {\enquote {\bibinfo {title} {Single-tone pulse sequences and robust two-tone shaped pulses for three silicon spin qubits with always-on exchange},}\ }\href {\doibase 10.1103/PhysRevB.103.235314} {\bibfield  {journal} {\bibinfo  {journal} {Phys. Rev. B}\ }\textbf {\bibinfo {volume} {103}},\ \bibinfo {pages} {235314} (\bibinfo {year} {2021})}\BibitemShut {NoStop}%
\bibitem [{\citenamefont {Ballance}\ \emph {et~al.}(2016)\citenamefont {Ballance}, \citenamefont {Harty}, \citenamefont {Linke}, \citenamefont {Sepiol},\ and\ \citenamefont {Lucas}}]{ballance2016high}%
  \BibitemOpen
  \bibfield  {author} {\bibinfo {author} {\bibfnamefont {C.~J.}\ \bibnamefont {Ballance}}, \bibinfo {author} {\bibfnamefont {T.~P.}\ \bibnamefont {Harty}}, \bibinfo {author} {\bibfnamefont {N.~M.}\ \bibnamefont {Linke}}, \bibinfo {author} {\bibfnamefont {M.~A.}\ \bibnamefont {Sepiol}}, \ and\ \bibinfo {author} {\bibfnamefont {D.~M.}\ \bibnamefont {Lucas}},\ }\bibfield  {title} {\enquote {\bibinfo {title} {High-fidelity quantum logic gates using trapped-ion hyperfine qubits},}\ }\href {\doibase 10.1103/PhysRevLett.117.060504} {\bibfield  {journal} {\bibinfo  {journal} {Phys. Rev. Lett.}\ }\textbf {\bibinfo {volume} {117}},\ \bibinfo {pages} {060504} (\bibinfo {year} {2016})}\BibitemShut {NoStop}%
\bibitem [{\citenamefont {Gaebler}\ \emph {et~al.}(2016)\citenamefont {Gaebler}, \citenamefont {Tan}, \citenamefont {Lin}, \citenamefont {Wan}, \citenamefont {Bowler}, \citenamefont {Keith}, \citenamefont {Glancy}, \citenamefont {Coakley}, \citenamefont {Knill}, \citenamefont {Leibfried},\ and\ \citenamefont {Wineland}}]{gaebler2016high}%
  \BibitemOpen
  \bibfield  {author} {\bibinfo {author} {\bibfnamefont {J.~P.}\ \bibnamefont {Gaebler}}, \bibinfo {author} {\bibfnamefont {T.~R.}\ \bibnamefont {Tan}}, \bibinfo {author} {\bibfnamefont {Y.}~\bibnamefont {Lin}}, \bibinfo {author} {\bibfnamefont {Y.}~\bibnamefont {Wan}}, \bibinfo {author} {\bibfnamefont {R.}~\bibnamefont {Bowler}}, \bibinfo {author} {\bibfnamefont {A.~C.}\ \bibnamefont {Keith}}, \bibinfo {author} {\bibfnamefont {S.}~\bibnamefont {Glancy}}, \bibinfo {author} {\bibfnamefont {K.}~\bibnamefont {Coakley}}, \bibinfo {author} {\bibfnamefont {E.}~\bibnamefont {Knill}}, \bibinfo {author} {\bibfnamefont {D.}~\bibnamefont {Leibfried}}, \ and\ \bibinfo {author} {\bibfnamefont {D.~J.}\ \bibnamefont {Wineland}},\ }\bibfield  {title} {\enquote {\bibinfo {title} {High-fidelity universal gate set for ${^{9}\mathrm{Be}}^{+}$ ion qubits},}\ }\href {\doibase 10.1103/PhysRevLett.117.060505} {\bibfield  {journal} {\bibinfo  {journal} {Phys. Rev. Lett.}\ }\textbf {\bibinfo {volume} {117}},\ \bibinfo {pages} {060505}
  (\bibinfo {year} {2016})}\BibitemShut {NoStop}%
\bibitem [{\citenamefont {Barends}\ \emph {et~al.}(2014)\citenamefont {Barends}, \citenamefont {Kelly}, \citenamefont {Megrant}, \citenamefont {Veitia}, \citenamefont {Sank}, \citenamefont {Jeffrey}, \citenamefont {White}, \citenamefont {Mutus}, \citenamefont {Fowler}, \citenamefont {Campbell} \emph {et~al.}}]{barends2014superconducting}%
  \BibitemOpen
  \bibfield  {author} {\bibinfo {author} {\bibfnamefont {Rami}\ \bibnamefont {Barends}}, \bibinfo {author} {\bibfnamefont {Julian}\ \bibnamefont {Kelly}}, \bibinfo {author} {\bibfnamefont {Anthony}\ \bibnamefont {Megrant}}, \bibinfo {author} {\bibfnamefont {Andrzej}\ \bibnamefont {Veitia}}, \bibinfo {author} {\bibfnamefont {Daniel}\ \bibnamefont {Sank}}, \bibinfo {author} {\bibfnamefont {Evan}\ \bibnamefont {Jeffrey}}, \bibinfo {author} {\bibfnamefont {Ted~C}\ \bibnamefont {White}}, \bibinfo {author} {\bibfnamefont {Josh}\ \bibnamefont {Mutus}}, \bibinfo {author} {\bibfnamefont {Austin~G}\ \bibnamefont {Fowler}}, \bibinfo {author} {\bibfnamefont {Brooks}\ \bibnamefont {Campbell}},  \emph {et~al.},\ }\bibfield  {title} {\enquote {\bibinfo {title} {Superconducting quantum circuits at the surface code threshold for fault tolerance},}\ }\href {https://doi.org/10.1038/nature13171} {\bibfield  {journal} {\bibinfo  {journal} {Nature}\ }\textbf {\bibinfo {volume} {508}},\ \bibinfo {pages} {500--503} (\bibinfo {year}
  {2014})}\BibitemShut {NoStop}%
\bibitem [{\citenamefont {Wang}\ \emph {et~al.}(2019)\citenamefont {Wang}, \citenamefont {Zhang}, \citenamefont {Xiang}, \citenamefont {Jia}, \citenamefont {Duan}, \citenamefont {Zong}, \citenamefont {Sun}, \citenamefont {Dong}, \citenamefont {Wu}, \citenamefont {Yin},\ and\ \citenamefont {Guo}}]{wang2019experimental}%
  \BibitemOpen
  \bibfield  {author} {\bibinfo {author} {\bibfnamefont {Tenghui}\ \bibnamefont {Wang}}, \bibinfo {author} {\bibfnamefont {Zhenxing}\ \bibnamefont {Zhang}}, \bibinfo {author} {\bibfnamefont {Liang}\ \bibnamefont {Xiang}}, \bibinfo {author} {\bibfnamefont {Zhilong}\ \bibnamefont {Jia}}, \bibinfo {author} {\bibfnamefont {Peng}\ \bibnamefont {Duan}}, \bibinfo {author} {\bibfnamefont {Zhiwen}\ \bibnamefont {Zong}}, \bibinfo {author} {\bibfnamefont {Zhenhai}\ \bibnamefont {Sun}}, \bibinfo {author} {\bibfnamefont {Zhangjingzi}\ \bibnamefont {Dong}}, \bibinfo {author} {\bibfnamefont {Jianlan}\ \bibnamefont {Wu}}, \bibinfo {author} {\bibfnamefont {Yi}~\bibnamefont {Yin}}, \ and\ \bibinfo {author} {\bibfnamefont {Guoping}\ \bibnamefont {Guo}},\ }\bibfield  {title} {\enquote {\bibinfo {title} {Experimental realization of a fast controlled-z gate via a shortcut to adiabaticity},}\ }\href {\doibase 10.1103/PhysRevApplied.11.034030} {\bibfield  {journal} {\bibinfo  {journal} {Phys. Rev. Appl.}\ }\textbf {\bibinfo {volume}
  {11}},\ \bibinfo {pages} {034030} (\bibinfo {year} {2019})}\BibitemShut {NoStop}%
\bibitem [{\citenamefont {Rol}\ \emph {et~al.}(2019)\citenamefont {Rol}, \citenamefont {Battistel}, \citenamefont {Malinowski}, \citenamefont {Bultink}, \citenamefont {Tarasinski}, \citenamefont {Vollmer}, \citenamefont {Haider}, \citenamefont {Muthusubramanian}, \citenamefont {Bruno}, \citenamefont {Terhal},\ and\ \citenamefont {DiCarlo}}]{rol2019fast}%
  \BibitemOpen
  \bibfield  {author} {\bibinfo {author} {\bibfnamefont {M.~A.}\ \bibnamefont {Rol}}, \bibinfo {author} {\bibfnamefont {F.}~\bibnamefont {Battistel}}, \bibinfo {author} {\bibfnamefont {F.~K.}\ \bibnamefont {Malinowski}}, \bibinfo {author} {\bibfnamefont {C.~C.}\ \bibnamefont {Bultink}}, \bibinfo {author} {\bibfnamefont {B.~M.}\ \bibnamefont {Tarasinski}}, \bibinfo {author} {\bibfnamefont {R.}~\bibnamefont {Vollmer}}, \bibinfo {author} {\bibfnamefont {N.}~\bibnamefont {Haider}}, \bibinfo {author} {\bibfnamefont {N.}~\bibnamefont {Muthusubramanian}}, \bibinfo {author} {\bibfnamefont {A.}~\bibnamefont {Bruno}}, \bibinfo {author} {\bibfnamefont {B.~M.}\ \bibnamefont {Terhal}}, \ and\ \bibinfo {author} {\bibfnamefont {L.}~\bibnamefont {DiCarlo}},\ }\bibfield  {title} {\enquote {\bibinfo {title} {Fast, high-fidelity conditional-phase gate exploiting leakage interference in weakly anharmonic superconducting qubits},}\ }\href {\doibase 10.1103/PhysRevLett.123.120502} {\bibfield  {journal} {\bibinfo  {journal} {Phys.
  Rev. Lett.}\ }\textbf {\bibinfo {volume} {123}},\ \bibinfo {pages} {120502} (\bibinfo {year} {2019})}\BibitemShut {NoStop}%
\bibitem [{\citenamefont {Lacroix}\ \emph {et~al.}(2020)\citenamefont {Lacroix}, \citenamefont {Hellings}, \citenamefont {Andersen}, \citenamefont {Di~Paolo}, \citenamefont {Remm}, \citenamefont {Lazar}, \citenamefont {Krinner}, \citenamefont {Norris}, \citenamefont {Gabureac}, \citenamefont {Heinsoo}, \citenamefont {Blais}, \citenamefont {Eichler},\ and\ \citenamefont {Wallraff}}]{improve}%
  \BibitemOpen
  \bibfield  {author} {\bibinfo {author} {\bibfnamefont {Nathan}\ \bibnamefont {Lacroix}}, \bibinfo {author} {\bibfnamefont {Christoph}\ \bibnamefont {Hellings}}, \bibinfo {author} {\bibfnamefont {Christian~Kraglund}\ \bibnamefont {Andersen}}, \bibinfo {author} {\bibfnamefont {Agustin}\ \bibnamefont {Di~Paolo}}, \bibinfo {author} {\bibfnamefont {Ants}\ \bibnamefont {Remm}}, \bibinfo {author} {\bibfnamefont {Stefania}\ \bibnamefont {Lazar}}, \bibinfo {author} {\bibfnamefont {Sebastian}\ \bibnamefont {Krinner}}, \bibinfo {author} {\bibfnamefont {Graham~J.}\ \bibnamefont {Norris}}, \bibinfo {author} {\bibfnamefont {Mihai}\ \bibnamefont {Gabureac}}, \bibinfo {author} {\bibfnamefont {Johannes}\ \bibnamefont {Heinsoo}}, \bibinfo {author} {\bibfnamefont {Alexandre}\ \bibnamefont {Blais}}, \bibinfo {author} {\bibfnamefont {Christopher}\ \bibnamefont {Eichler}}, \ and\ \bibinfo {author} {\bibfnamefont {Andreas}\ \bibnamefont {Wallraff}},\ }\bibfield  {title} {\enquote {\bibinfo {title} {Improving the performance of
  deep quantum optimization algorithms with continuous gate sets},}\ }\href {\doibase 10.1103/PRXQuantum.1.020304} {\bibfield  {journal} {\bibinfo  {journal} {PRX Quantum}\ }\textbf {\bibinfo {volume} {1}},\ \bibinfo {pages} {020304} (\bibinfo {year} {2020})}\BibitemShut {NoStop}%
\bibitem [{\citenamefont {Bloch}(2008)}]{bloch2008quantum}%
  \BibitemOpen
  \bibfield  {author} {\bibinfo {author} {\bibfnamefont {Immanuel}\ \bibnamefont {Bloch}},\ }\bibfield  {title} {\enquote {\bibinfo {title} {Quantum coherence and entanglement with ultracold atoms in optical lattices},}\ }\href {https://doi.org/10.1038/nature07126} {\bibfield  {journal} {\bibinfo  {journal} {Nature}\ }\textbf {\bibinfo {volume} {453}},\ \bibinfo {pages} {1016--1022} (\bibinfo {year} {2008})}\BibitemShut {NoStop}%
\bibitem [{\citenamefont {Saffman}\ \emph {et~al.}(2010)\citenamefont {Saffman}, \citenamefont {Walker},\ and\ \citenamefont {M\o{}lmer}}]{QuantumInformationWithRydbergAtoms}%
  \BibitemOpen
  \bibfield  {author} {\bibinfo {author} {\bibfnamefont {M.}~\bibnamefont {Saffman}}, \bibinfo {author} {\bibfnamefont {T.~G.}\ \bibnamefont {Walker}}, \ and\ \bibinfo {author} {\bibfnamefont {K.}~\bibnamefont {M\o{}lmer}},\ }\bibfield  {title} {\enquote {\bibinfo {title} {Quantum information with rydberg atoms},}\ }\href {\doibase 10.1103/RevModPhys.82.2313} {\bibfield  {journal} {\bibinfo  {journal} {Rev. Mod. Phys.}\ }\textbf {\bibinfo {volume} {82}},\ \bibinfo {pages} {2313--2363} (\bibinfo {year} {2010})}\BibitemShut {NoStop}%
\bibitem [{\citenamefont {Browaeys}\ \emph {et~al.}(2016)\citenamefont {Browaeys}, \citenamefont {Barredo},\ and\ \citenamefont {Lahaye}}]{browaeys2016experimental}%
  \BibitemOpen
  \bibfield  {author} {\bibinfo {author} {\bibfnamefont {Antoine}\ \bibnamefont {Browaeys}}, \bibinfo {author} {\bibfnamefont {Daniel}\ \bibnamefont {Barredo}}, \ and\ \bibinfo {author} {\bibfnamefont {Thierry}\ \bibnamefont {Lahaye}},\ }\bibfield  {title} {\enquote {\bibinfo {title} {Experimental investigations of dipole--dipole interactions between a few rydberg atoms},}\ }\href {https://iopscience.iop.org/article/10.1088/0953-4075/49/15/152001/meta} {\bibfield  {journal} {\bibinfo  {journal} {Journal of Physics B: Atomic, Molecular and Optical Physics}\ }\textbf {\bibinfo {volume} {49}},\ \bibinfo {pages} {152001} (\bibinfo {year} {2016})}\BibitemShut {NoStop}%
\bibitem [{\citenamefont {Shao}\ \emph {et~al.}(2024)\citenamefont {Shao}, \citenamefont {Su}, \citenamefont {Li}, \citenamefont {Nath}, \citenamefont {Wu},\ and\ \citenamefont {Li}}]{shao2024rydberg}%
  \BibitemOpen
  \bibfield  {author} {\bibinfo {author} {\bibfnamefont {Xiao-Qiang}\ \bibnamefont {Shao}}, \bibinfo {author} {\bibfnamefont {Shi-Lei}\ \bibnamefont {Su}}, \bibinfo {author} {\bibfnamefont {Lin}\ \bibnamefont {Li}}, \bibinfo {author} {\bibfnamefont {Rejish}\ \bibnamefont {Nath}}, \bibinfo {author} {\bibfnamefont {Jin-Hui}\ \bibnamefont {Wu}}, \ and\ \bibinfo {author} {\bibfnamefont {Weibin}\ \bibnamefont {Li}},\ }\bibfield  {title} {\enquote {\bibinfo {title} {Rydberg superatoms: An artificial quantum system for quantum information processing and quantum optics},}\ }\href {https://doi.org/10.48550/arXiv.2404.05330} {\bibfield  {journal} {\bibinfo  {journal} {arXiv preprint arXiv:2404.05330}\ } (\bibinfo {year} {2024})}\BibitemShut {NoStop}%
\bibitem [{\citenamefont {Urban}\ \emph {et~al.}(2009)\citenamefont {Urban}, \citenamefont {Johnson}, \citenamefont {Henage}, \citenamefont {Isenhower}, \citenamefont {Yavuz}, \citenamefont {Walker},\ and\ \citenamefont {Saffman}}]{urban2009observation}%
  \BibitemOpen
  \bibfield  {author} {\bibinfo {author} {\bibfnamefont {E}~\bibnamefont {Urban}}, \bibinfo {author} {\bibfnamefont {Todd~A}\ \bibnamefont {Johnson}}, \bibinfo {author} {\bibfnamefont {T}~\bibnamefont {Henage}}, \bibinfo {author} {\bibfnamefont {L}~\bibnamefont {Isenhower}}, \bibinfo {author} {\bibfnamefont {DD}~\bibnamefont {Yavuz}}, \bibinfo {author} {\bibfnamefont {TG}~\bibnamefont {Walker}}, \ and\ \bibinfo {author} {\bibfnamefont {M}~\bibnamefont {Saffman}},\ }\bibfield  {title} {\enquote {\bibinfo {title} {Observation of rydberg blockade between two atoms},}\ }\href {https://doi.org/10.1038/nphys1178} {\bibfield  {journal} {\bibinfo  {journal} {Nature Physics}\ }\textbf {\bibinfo {volume} {5}},\ \bibinfo {pages} {110--114} (\bibinfo {year} {2009})}\BibitemShut {NoStop}%
\bibitem [{\citenamefont {Tong}\ \emph {et~al.}(2004)\citenamefont {Tong}, \citenamefont {Farooqi}, \citenamefont {Stanojevic}, \citenamefont {Krishnan}, \citenamefont {Zhang}, \citenamefont {C\^ot\'e}, \citenamefont {Eyler},\ and\ \citenamefont {Gould}}]{tong2004local}%
  \BibitemOpen
  \bibfield  {author} {\bibinfo {author} {\bibfnamefont {D.}~\bibnamefont {Tong}}, \bibinfo {author} {\bibfnamefont {S.~M.}\ \bibnamefont {Farooqi}}, \bibinfo {author} {\bibfnamefont {J.}~\bibnamefont {Stanojevic}}, \bibinfo {author} {\bibfnamefont {S.}~\bibnamefont {Krishnan}}, \bibinfo {author} {\bibfnamefont {Y.~P.}\ \bibnamefont {Zhang}}, \bibinfo {author} {\bibfnamefont {R.}~\bibnamefont {C\^ot\'e}}, \bibinfo {author} {\bibfnamefont {E.~E.}\ \bibnamefont {Eyler}}, \ and\ \bibinfo {author} {\bibfnamefont {P.~L.}\ \bibnamefont {Gould}},\ }\bibfield  {title} {\enquote {\bibinfo {title} {Local blockade of rydberg excitation in an ultracold gas},}\ }\href {\doibase 10.1103/PhysRevLett.93.063001} {\bibfield  {journal} {\bibinfo  {journal} {Phys. Rev. Lett.}\ }\textbf {\bibinfo {volume} {93}},\ \bibinfo {pages} {063001} (\bibinfo {year} {2004})}\BibitemShut {NoStop}%
\bibitem [{\citenamefont {Ga{\"e}tan}\ \emph {et~al.}(2009)\citenamefont {Ga{\"e}tan}, \citenamefont {Miroshnychenko}, \citenamefont {Wilk}, \citenamefont {Chotia}, \citenamefont {Viteau}, \citenamefont {Comparat}, \citenamefont {Pillet}, \citenamefont {Browaeys},\ and\ \citenamefont {Grangier}}]{gaetan2009observation}%
  \BibitemOpen
  \bibfield  {author} {\bibinfo {author} {\bibfnamefont {Alpha}\ \bibnamefont {Ga{\"e}tan}}, \bibinfo {author} {\bibfnamefont {Yevhen}\ \bibnamefont {Miroshnychenko}}, \bibinfo {author} {\bibfnamefont {Tatjana}\ \bibnamefont {Wilk}}, \bibinfo {author} {\bibfnamefont {Amodsen}\ \bibnamefont {Chotia}}, \bibinfo {author} {\bibfnamefont {Matthieu}\ \bibnamefont {Viteau}}, \bibinfo {author} {\bibfnamefont {Daniel}\ \bibnamefont {Comparat}}, \bibinfo {author} {\bibfnamefont {Pierre}\ \bibnamefont {Pillet}}, \bibinfo {author} {\bibfnamefont {Antoine}\ \bibnamefont {Browaeys}}, \ and\ \bibinfo {author} {\bibfnamefont {Philippe}\ \bibnamefont {Grangier}},\ }\bibfield  {title} {\enquote {\bibinfo {title} {Observation of collective excitation of two individual atoms in the rydberg blockade regime},}\ }\href {https://doi.org/10.1038/nphys1183} {\bibfield  {journal} {\bibinfo  {journal} {Nature Physics}\ }\textbf {\bibinfo {volume} {5}},\ \bibinfo {pages} {115--118} (\bibinfo {year} {2009})}\BibitemShut {NoStop}%
\bibitem [{\citenamefont {Jaksch}\ \emph {et~al.}(2000)\citenamefont {Jaksch}, \citenamefont {Cirac}, \citenamefont {Zoller}, \citenamefont {Rolston}, \citenamefont {C\^ot\'e},\ and\ \citenamefont {Lukin}}]{jaksch2000fast}%
  \BibitemOpen
  \bibfield  {author} {\bibinfo {author} {\bibfnamefont {D.}~\bibnamefont {Jaksch}}, \bibinfo {author} {\bibfnamefont {J.~I.}\ \bibnamefont {Cirac}}, \bibinfo {author} {\bibfnamefont {P.}~\bibnamefont {Zoller}}, \bibinfo {author} {\bibfnamefont {S.~L.}\ \bibnamefont {Rolston}}, \bibinfo {author} {\bibfnamefont {R.}~\bibnamefont {C\^ot\'e}}, \ and\ \bibinfo {author} {\bibfnamefont {M.~D.}\ \bibnamefont {Lukin}},\ }\bibfield  {title} {\enquote {\bibinfo {title} {Fast quantum gates for neutral atoms},}\ }\href {\doibase 10.1103/PhysRevLett.85.2208} {\bibfield  {journal} {\bibinfo  {journal} {Phys. Rev. Lett.}\ }\textbf {\bibinfo {volume} {85}},\ \bibinfo {pages} {2208--2211} (\bibinfo {year} {2000})}\BibitemShut {NoStop}%
\bibitem [{\citenamefont {Lukin}\ \emph {et~al.}(2001)\citenamefont {Lukin}, \citenamefont {Fleischhauer}, \citenamefont {Cote}, \citenamefont {Duan}, \citenamefont {Jaksch}, \citenamefont {Cirac},\ and\ \citenamefont {Zoller}}]{lukin2001dipole}%
  \BibitemOpen
  \bibfield  {author} {\bibinfo {author} {\bibfnamefont {M.~D.}\ \bibnamefont {Lukin}}, \bibinfo {author} {\bibfnamefont {M.}~\bibnamefont {Fleischhauer}}, \bibinfo {author} {\bibfnamefont {R.}~\bibnamefont {Cote}}, \bibinfo {author} {\bibfnamefont {L.~M.}\ \bibnamefont {Duan}}, \bibinfo {author} {\bibfnamefont {D.}~\bibnamefont {Jaksch}}, \bibinfo {author} {\bibfnamefont {J.~I.}\ \bibnamefont {Cirac}}, \ and\ \bibinfo {author} {\bibfnamefont {P.}~\bibnamefont {Zoller}},\ }\bibfield  {title} {\enquote {\bibinfo {title} {Dipole blockade and quantum information processing in mesoscopic atomic ensembles},}\ }\href {\doibase 10.1103/PhysRevLett.87.037901} {\bibfield  {journal} {\bibinfo  {journal} {Phys. Rev. Lett.}\ }\textbf {\bibinfo {volume} {87}},\ \bibinfo {pages} {037901} (\bibinfo {year} {2001})}\BibitemShut {NoStop}%
\bibitem [{\citenamefont {Brion}\ \emph {et~al.}(2007)\citenamefont {Brion}, \citenamefont {Pedersen},\ and\ \citenamefont {M{\o}lmer}}]{brion2007implementing}%
  \BibitemOpen
  \bibfield  {author} {\bibinfo {author} {\bibfnamefont {E}~\bibnamefont {Brion}}, \bibinfo {author} {\bibfnamefont {LH}~\bibnamefont {Pedersen}}, \ and\ \bibinfo {author} {\bibfnamefont {K}~\bibnamefont {M{\o}lmer}},\ }\bibfield  {title} {\enquote {\bibinfo {title} {Implementing a neutral atom rydberg gate without populating the rydberg state},}\ }\href {\doibase 10.1088/0953-4075/40/9/S09} {\bibfield  {journal} {\bibinfo  {journal} {Journal of Physics B: Atomic, Molecular and Optical Physics}\ }\textbf {\bibinfo {volume} {40}},\ \bibinfo {pages} {S159} (\bibinfo {year} {2007})}\BibitemShut {NoStop}%
\bibitem [{\citenamefont {Wu}\ \emph {et~al.}(2010)\citenamefont {Wu}, \citenamefont {Yang},\ and\ \citenamefont {Zheng}}]{wu2010implementation}%
  \BibitemOpen
  \bibfield  {author} {\bibinfo {author} {\bibfnamefont {Huai-Zhi}\ \bibnamefont {Wu}}, \bibinfo {author} {\bibfnamefont {Zhen-Biao}\ \bibnamefont {Yang}}, \ and\ \bibinfo {author} {\bibfnamefont {Shi-Biao}\ \bibnamefont {Zheng}},\ }\bibfield  {title} {\enquote {\bibinfo {title} {Implementation of a multiqubit quantum phase gate in a neutral atomic ensemble via the asymmetric rydberg blockade},}\ }\href {\doibase 10.1103/PhysRevA.82.034307} {\bibfield  {journal} {\bibinfo  {journal} {Phys. Rev. A}\ }\textbf {\bibinfo {volume} {82}},\ \bibinfo {pages} {034307} (\bibinfo {year} {2010})}\BibitemShut {NoStop}%
\bibitem [{\citenamefont {M\"uller}\ \emph {et~al.}(2011)\citenamefont {M\"uller}, \citenamefont {Reich}, \citenamefont {Murphy}, \citenamefont {Yuan}, \citenamefont {Vala}, \citenamefont {Whaley}, \citenamefont {Calarco},\ and\ \citenamefont {Koch}}]{muller2011optimizing}%
  \BibitemOpen
  \bibfield  {author} {\bibinfo {author} {\bibfnamefont {M.~M.}\ \bibnamefont {M\"uller}}, \bibinfo {author} {\bibfnamefont {D.~M.}\ \bibnamefont {Reich}}, \bibinfo {author} {\bibfnamefont {M.}~\bibnamefont {Murphy}}, \bibinfo {author} {\bibfnamefont {H.}~\bibnamefont {Yuan}}, \bibinfo {author} {\bibfnamefont {J.}~\bibnamefont {Vala}}, \bibinfo {author} {\bibfnamefont {K.~B.}\ \bibnamefont {Whaley}}, \bibinfo {author} {\bibfnamefont {T.}~\bibnamefont {Calarco}}, \ and\ \bibinfo {author} {\bibfnamefont {C.~P.}\ \bibnamefont {Koch}},\ }\bibfield  {title} {\enquote {\bibinfo {title} {Optimizing entangling quantum gates for physical systems},}\ }\href {\doibase 10.1103/PhysRevA.84.042315} {\bibfield  {journal} {\bibinfo  {journal} {Phys. Rev. A}\ }\textbf {\bibinfo {volume} {84}},\ \bibinfo {pages} {042315} (\bibinfo {year} {2011})}\BibitemShut {NoStop}%
\bibitem [{\citenamefont {Xia}\ \emph {et~al.}(2013)\citenamefont {Xia}, \citenamefont {Zhang},\ and\ \citenamefont {Saffman}}]{xia2013analysis}%
  \BibitemOpen
  \bibfield  {author} {\bibinfo {author} {\bibfnamefont {T.}~\bibnamefont {Xia}}, \bibinfo {author} {\bibfnamefont {X.~L.}\ \bibnamefont {Zhang}}, \ and\ \bibinfo {author} {\bibfnamefont {M.}~\bibnamefont {Saffman}},\ }\bibfield  {title} {\enquote {\bibinfo {title} {Analysis of a controlled phase gate using circular rydberg states},}\ }\href {\doibase 10.1103/PhysRevA.88.062337} {\bibfield  {journal} {\bibinfo  {journal} {Phys. Rev. A}\ }\textbf {\bibinfo {volume} {88}},\ \bibinfo {pages} {062337} (\bibinfo {year} {2013})}\BibitemShut {NoStop}%
\bibitem [{\citenamefont {Petrosyan}\ and\ \citenamefont {M\o{}lmer}(2014)}]{petrosyan2014binding}%
  \BibitemOpen
  \bibfield  {author} {\bibinfo {author} {\bibfnamefont {David}\ \bibnamefont {Petrosyan}}\ and\ \bibinfo {author} {\bibfnamefont {Klaus}\ \bibnamefont {M\o{}lmer}},\ }\bibfield  {title} {\enquote {\bibinfo {title} {Binding potentials and interaction gates between microwave-dressed rydberg atoms},}\ }\href {\doibase 10.1103/PhysRevLett.113.123003} {\bibfield  {journal} {\bibinfo  {journal} {Phys. Rev. Lett.}\ }\textbf {\bibinfo {volume} {113}},\ \bibinfo {pages} {123003} (\bibinfo {year} {2014})}\BibitemShut {NoStop}%
\bibitem [{\citenamefont {S\'ark\'any}\ \emph {et~al.}(2015)\citenamefont {S\'ark\'any}, \citenamefont {Fort\'agh},\ and\ \citenamefont {Petrosyan}}]{sarkany2015long}%
  \BibitemOpen
  \bibfield  {author} {\bibinfo {author} {\bibfnamefont {L\"orinc}\ \bibnamefont {S\'ark\'any}}, \bibinfo {author} {\bibfnamefont {J\'ozsef}\ \bibnamefont {Fort\'agh}}, \ and\ \bibinfo {author} {\bibfnamefont {David}\ \bibnamefont {Petrosyan}},\ }\bibfield  {title} {\enquote {\bibinfo {title} {Long-range quantum gate via rydberg states of atoms in a thermal microwave cavity},}\ }\href {\doibase 10.1103/PhysRevA.92.030303} {\bibfield  {journal} {\bibinfo  {journal} {Phys. Rev. A}\ }\textbf {\bibinfo {volume} {92}},\ \bibinfo {pages} {030303} (\bibinfo {year} {2015})}\BibitemShut {NoStop}%
\bibitem [{\citenamefont {Saffman}(2016)}]{saffman2016quantum}%
  \BibitemOpen
  \bibfield  {author} {\bibinfo {author} {\bibfnamefont {Mark}\ \bibnamefont {Saffman}},\ }\bibfield  {title} {\enquote {\bibinfo {title} {Quantum computing with atomic qubits and rydberg interactions: progress and challenges},}\ }\href {\doibase 10.1088/0953-4075/49/20/202001} {\bibfield  {journal} {\bibinfo  {journal} {Journal of Physics B: Atomic, Molecular and Optical Physics}\ }\textbf {\bibinfo {volume} {49}},\ \bibinfo {pages} {202001} (\bibinfo {year} {2016})}\BibitemShut {NoStop}%
\bibitem [{\citenamefont {Su}\ \emph {et~al.}(2016)\citenamefont {Su}, \citenamefont {Liang}, \citenamefont {Zhang}, \citenamefont {Wen}, \citenamefont {Sun}, \citenamefont {Jin},\ and\ \citenamefont {Zhu}}]{su2016one}%
  \BibitemOpen
  \bibfield  {author} {\bibinfo {author} {\bibfnamefont {Shi-Lei}\ \bibnamefont {Su}}, \bibinfo {author} {\bibfnamefont {Erjun}\ \bibnamefont {Liang}}, \bibinfo {author} {\bibfnamefont {Shou}\ \bibnamefont {Zhang}}, \bibinfo {author} {\bibfnamefont {Jing-Ji}\ \bibnamefont {Wen}}, \bibinfo {author} {\bibfnamefont {Li-Li}\ \bibnamefont {Sun}}, \bibinfo {author} {\bibfnamefont {Zhao}\ \bibnamefont {Jin}}, \ and\ \bibinfo {author} {\bibfnamefont {Ai-Dong}\ \bibnamefont {Zhu}},\ }\bibfield  {title} {\enquote {\bibinfo {title} {One-step implementation of the rydberg-rydberg-interaction gate},}\ }\href {\doibase 10.1103/PhysRevA.93.012306} {\bibfield  {journal} {\bibinfo  {journal} {Phys. Rev. A}\ }\textbf {\bibinfo {volume} {93}},\ \bibinfo {pages} {012306} (\bibinfo {year} {2016})}\BibitemShut {NoStop}%
\bibitem [{\citenamefont {Su}\ \emph {et~al.}(2017)\citenamefont {Su}, \citenamefont {Gao}, \citenamefont {Liang},\ and\ \citenamefont {Zhang}}]{su2017fast}%
  \BibitemOpen
  \bibfield  {author} {\bibinfo {author} {\bibfnamefont {Shi-Lei}\ \bibnamefont {Su}}, \bibinfo {author} {\bibfnamefont {Ya}~\bibnamefont {Gao}}, \bibinfo {author} {\bibfnamefont {Erjun}\ \bibnamefont {Liang}}, \ and\ \bibinfo {author} {\bibfnamefont {Shou}\ \bibnamefont {Zhang}},\ }\bibfield  {title} {\enquote {\bibinfo {title} {Fast rydberg antiblockade regime and its applications in quantum logic gates},}\ }\href {\doibase 10.1103/PhysRevA.95.022319} {\bibfield  {journal} {\bibinfo  {journal} {Phys. Rev. A}\ }\textbf {\bibinfo {volume} {95}},\ \bibinfo {pages} {022319} (\bibinfo {year} {2017})}\BibitemShut {NoStop}%
\bibitem [{\citenamefont {Shi}(2018)}]{shi2018deutsch}%
  \BibitemOpen
  \bibfield  {author} {\bibinfo {author} {\bibfnamefont {Xiao-Feng}\ \bibnamefont {Shi}},\ }\bibfield  {title} {\enquote {\bibinfo {title} {Deutsch, toffoli, and cnot gates via rydberg blockade of neutral atoms},}\ }\href {\doibase 10.1103/PhysRevApplied.9.051001} {\bibfield  {journal} {\bibinfo  {journal} {Phys. Rev. Appl.}\ }\textbf {\bibinfo {volume} {9}},\ \bibinfo {pages} {051001} (\bibinfo {year} {2018})}\BibitemShut {NoStop}%
\bibitem [{\citenamefont {Shi}\ and\ \citenamefont {Kennedy}(2017)}]{shi2017annulled}%
  \BibitemOpen
  \bibfield  {author} {\bibinfo {author} {\bibfnamefont {Xiao-Feng}\ \bibnamefont {Shi}}\ and\ \bibinfo {author} {\bibfnamefont {T.~A.~B.}\ \bibnamefont {Kennedy}},\ }\bibfield  {title} {\enquote {\bibinfo {title} {Annulled van der waals interaction and fast rydberg quantum gates},}\ }\href {\doibase 10.1103/PhysRevA.95.043429} {\bibfield  {journal} {\bibinfo  {journal} {Phys. Rev. A}\ }\textbf {\bibinfo {volume} {95}},\ \bibinfo {pages} {043429} (\bibinfo {year} {2017})}\BibitemShut {NoStop}%
\bibitem [{\citenamefont {Shi}(2017)}]{shi2017rydberg}%
  \BibitemOpen
  \bibfield  {author} {\bibinfo {author} {\bibfnamefont {Xiao-Feng}\ \bibnamefont {Shi}},\ }\bibfield  {title} {\enquote {\bibinfo {title} {Rydberg quantum gates free from blockade error},}\ }\href {\doibase 10.1103/PhysRevApplied.7.064017} {\bibfield  {journal} {\bibinfo  {journal} {Phys. Rev. Appl.}\ }\textbf {\bibinfo {volume} {7}},\ \bibinfo {pages} {064017} (\bibinfo {year} {2017})}\BibitemShut {NoStop}%
\bibitem [{\citenamefont {Huang}\ \emph {et~al.}(2018)\citenamefont {Huang}, \citenamefont {Ding}, \citenamefont {Hu}, \citenamefont {Shen}, \citenamefont {Li}, \citenamefont {Wu},\ and\ \citenamefont {Zheng}}]{huang2018robust}%
  \BibitemOpen
  \bibfield  {author} {\bibinfo {author} {\bibfnamefont {Xi-Rong}\ \bibnamefont {Huang}}, \bibinfo {author} {\bibfnamefont {Zong-Xing}\ \bibnamefont {Ding}}, \bibinfo {author} {\bibfnamefont {Chang-Sheng}\ \bibnamefont {Hu}}, \bibinfo {author} {\bibfnamefont {Li-Tuo}\ \bibnamefont {Shen}}, \bibinfo {author} {\bibfnamefont {Weibin}\ \bibnamefont {Li}}, \bibinfo {author} {\bibfnamefont {Huaizhi}\ \bibnamefont {Wu}}, \ and\ \bibinfo {author} {\bibfnamefont {Shi-Biao}\ \bibnamefont {Zheng}},\ }\bibfield  {title} {\enquote {\bibinfo {title} {Robust rydberg gate via landau-zener control of f\"orster resonance},}\ }\href {\doibase 10.1103/PhysRevA.98.052324} {\bibfield  {journal} {\bibinfo  {journal} {Phys. Rev. A}\ }\textbf {\bibinfo {volume} {98}},\ \bibinfo {pages} {052324} (\bibinfo {year} {2018})}\BibitemShut {NoStop}%
\bibitem [{\citenamefont {Su}\ \emph {et~al.}(2018)\citenamefont {Su}, \citenamefont {Shen}, \citenamefont {Liang},\ and\ \citenamefont {Zhang}}]{su2018one}%
  \BibitemOpen
  \bibfield  {author} {\bibinfo {author} {\bibfnamefont {S.~L.}\ \bibnamefont {Su}}, \bibinfo {author} {\bibfnamefont {H.~Z.}\ \bibnamefont {Shen}}, \bibinfo {author} {\bibfnamefont {Erjun}\ \bibnamefont {Liang}}, \ and\ \bibinfo {author} {\bibfnamefont {Shou}\ \bibnamefont {Zhang}},\ }\bibfield  {title} {\enquote {\bibinfo {title} {One-step construction of the multiple-qubit rydberg controlled-phase gate},}\ }\href {\doibase 10.1103/PhysRevA.98.032306} {\bibfield  {journal} {\bibinfo  {journal} {Phys. Rev. A}\ }\textbf {\bibinfo {volume} {98}},\ \bibinfo {pages} {032306} (\bibinfo {year} {2018})}\BibitemShut {NoStop}%
\bibitem [{\citenamefont {Li}\ and\ \citenamefont {Shao}(2018)}]{li2018unconventional}%
  \BibitemOpen
  \bibfield  {author} {\bibinfo {author} {\bibfnamefont {D.~X.}\ \bibnamefont {Li}}\ and\ \bibinfo {author} {\bibfnamefont {X.~Q.}\ \bibnamefont {Shao}},\ }\bibfield  {title} {\enquote {\bibinfo {title} {Unconventional rydberg pumping and applications in quantum information processing},}\ }\href {\doibase 10.1103/PhysRevA.98.062338} {\bibfield  {journal} {\bibinfo  {journal} {Phys. Rev. A}\ }\textbf {\bibinfo {volume} {98}},\ \bibinfo {pages} {062338} (\bibinfo {year} {2018})}\BibitemShut {NoStop}%
\bibitem [{\citenamefont {Shi}(2019)}]{shi2019fast}%
  \BibitemOpen
  \bibfield  {author} {\bibinfo {author} {\bibfnamefont {Xiao-Feng}\ \bibnamefont {Shi}},\ }\bibfield  {title} {\enquote {\bibinfo {title} {Fast, accurate, and realizable two-qubit entangling gates by quantum interference in detuned rabi cycles of rydberg atoms},}\ }\href {\doibase 10.1103/PhysRevApplied.11.044035} {\bibfield  {journal} {\bibinfo  {journal} {Phys. Rev. Appl.}\ }\textbf {\bibinfo {volume} {11}},\ \bibinfo {pages} {044035} (\bibinfo {year} {2019})}\BibitemShut {NoStop}%
\bibitem [{\citenamefont {Yin}\ \emph {et~al.}(2020)\citenamefont {Yin}, \citenamefont {Li}, \citenamefont {Wang},\ and\ \citenamefont {Shao}}]{yin2020one}%
  \BibitemOpen
  \bibfield  {author} {\bibinfo {author} {\bibfnamefont {Hong-Da}\ \bibnamefont {Yin}}, \bibinfo {author} {\bibfnamefont {Xiao-Xuan}\ \bibnamefont {Li}}, \bibinfo {author} {\bibfnamefont {Gang-Cheng}\ \bibnamefont {Wang}}, \ and\ \bibinfo {author} {\bibfnamefont {Xiao-Qiang}\ \bibnamefont {Shao}},\ }\bibfield  {title} {\enquote {\bibinfo {title} {One-step implementation of toffoli gate for neutral atoms based on unconventional rydberg pumping},}\ }\href {\doibase 10.1364/OE.410158} {\bibfield  {journal} {\bibinfo  {journal} {Opt. Express}\ }\textbf {\bibinfo {volume} {28}},\ \bibinfo {pages} {35576--35587} (\bibinfo {year} {2020})}\BibitemShut {NoStop}%
\bibitem [{\citenamefont {Li}\ \emph {et~al.}(2022{\natexlab{a}})\citenamefont {Li}, \citenamefont {Li}, \citenamefont {Yu}, \citenamefont {Qian},\ and\ \citenamefont {Zhang}}]{li2022optimal}%
  \BibitemOpen
  \bibfield  {author} {\bibinfo {author} {\bibfnamefont {Rui}\ \bibnamefont {Li}}, \bibinfo {author} {\bibfnamefont {Shurui}\ \bibnamefont {Li}}, \bibinfo {author} {\bibfnamefont {Dongmin}\ \bibnamefont {Yu}}, \bibinfo {author} {\bibfnamefont {Jing}\ \bibnamefont {Qian}}, \ and\ \bibinfo {author} {\bibfnamefont {Weiping}\ \bibnamefont {Zhang}},\ }\bibfield  {title} {\enquote {\bibinfo {title} {Optimal model for fewer-qubit cnot gates with rydberg atoms},}\ }\href {\doibase 10.1103/PhysRevApplied.17.024014} {\bibfield  {journal} {\bibinfo  {journal} {Phys. Rev. Appl.}\ }\textbf {\bibinfo {volume} {17}},\ \bibinfo {pages} {024014} (\bibinfo {year} {2022}{\natexlab{a}})}\BibitemShut {NoStop}%
\bibitem [{\citenamefont {Li}\ \emph {et~al.}(2022{\natexlab{b}})\citenamefont {Li}, \citenamefont {You}, \citenamefont {Shao},\ and\ \citenamefont {Li}}]{li2022coherent}%
  \BibitemOpen
  \bibfield  {author} {\bibinfo {author} {\bibfnamefont {X.~X.}\ \bibnamefont {Li}}, \bibinfo {author} {\bibfnamefont {J.~B.}\ \bibnamefont {You}}, \bibinfo {author} {\bibfnamefont {X.~Q.}\ \bibnamefont {Shao}}, \ and\ \bibinfo {author} {\bibfnamefont {Weibin}\ \bibnamefont {Li}},\ }\bibfield  {title} {\enquote {\bibinfo {title} {Coherent ground-state transport of neutral atoms},}\ }\href {\doibase 10.1103/PhysRevA.105.032417} {\bibfield  {journal} {\bibinfo  {journal} {Phys. Rev. A}\ }\textbf {\bibinfo {volume} {105}},\ \bibinfo {pages} {032417} (\bibinfo {year} {2022}{\natexlab{b}})}\BibitemShut {NoStop}%
\bibitem [{\citenamefont {Levine}\ \emph {et~al.}(2019)\citenamefont {Levine}, \citenamefont {Keesling}, \citenamefont {Semeghini}, \citenamefont {Omran}, \citenamefont {Wang}, \citenamefont {Ebadi}, \citenamefont {Bernien}, \citenamefont {Greiner}, \citenamefont {Vuleti\ifmmode~\acute{c}\else \'{c}\fi{}}, \citenamefont {Pichler},\ and\ \citenamefont {Lukin}}]{levine2019parallel}%
  \BibitemOpen
  \bibfield  {author} {\bibinfo {author} {\bibfnamefont {Harry}\ \bibnamefont {Levine}}, \bibinfo {author} {\bibfnamefont {Alexander}\ \bibnamefont {Keesling}}, \bibinfo {author} {\bibfnamefont {Giulia}\ \bibnamefont {Semeghini}}, \bibinfo {author} {\bibfnamefont {Ahmed}\ \bibnamefont {Omran}}, \bibinfo {author} {\bibfnamefont {Tout~T.}\ \bibnamefont {Wang}}, \bibinfo {author} {\bibfnamefont {Sepehr}\ \bibnamefont {Ebadi}}, \bibinfo {author} {\bibfnamefont {Hannes}\ \bibnamefont {Bernien}}, \bibinfo {author} {\bibfnamefont {Markus}\ \bibnamefont {Greiner}}, \bibinfo {author} {\bibfnamefont {Vladan}\ \bibnamefont {Vuleti\ifmmode~\acute{c}\else \'{c}\fi{}}}, \bibinfo {author} {\bibfnamefont {Hannes}\ \bibnamefont {Pichler}}, \ and\ \bibinfo {author} {\bibfnamefont {Mikhail~D.}\ \bibnamefont {Lukin}},\ }\bibfield  {title} {\enquote {\bibinfo {title} {Parallel implementation of high-fidelity multiqubit gates with neutral atoms},}\ }\href {\doibase 10.1103/PhysRevLett.123.170503} {\bibfield  {journal} {\bibinfo
  {journal} {Phys. Rev. Lett.}\ }\textbf {\bibinfo {volume} {123}},\ \bibinfo {pages} {170503} (\bibinfo {year} {2019})}\BibitemShut {NoStop}%
\bibitem [{\citenamefont {Martin}\ \emph {et~al.}(2021)\citenamefont {Martin}, \citenamefont {Jau}, \citenamefont {Lee}, \citenamefont {Mitra}, \citenamefont {Deutsch},\ and\ \citenamefont {Biedermann}}]{martin2021molmer}%
  \BibitemOpen
  \bibfield  {author} {\bibinfo {author} {\bibfnamefont {Michael~J}\ \bibnamefont {Martin}}, \bibinfo {author} {\bibfnamefont {Yuan-Yu}\ \bibnamefont {Jau}}, \bibinfo {author} {\bibfnamefont {Jongmin}\ \bibnamefont {Lee}}, \bibinfo {author} {\bibfnamefont {Anupam}\ \bibnamefont {Mitra}}, \bibinfo {author} {\bibfnamefont {Ivan~H}\ \bibnamefont {Deutsch}}, \ and\ \bibinfo {author} {\bibfnamefont {Grant~W}\ \bibnamefont {Biedermann}},\ }\bibfield  {title} {\enquote {\bibinfo {title} {A m{\o}lmer-s{\o}rensen gate with rydberg-dressed atoms},}\ }\href {\doibase 10.48550/arXiv.2111.14677} {\bibfield  {journal} {\bibinfo  {journal} {arXiv e-prints}\ ,\ \bibinfo {pages} {arXiv--2111}} (\bibinfo {year} {2021})}\BibitemShut {NoStop}%
\bibitem [{\citenamefont {Fu}\ \emph {et~al.}(2022)\citenamefont {Fu}, \citenamefont {Xu}, \citenamefont {Sun}, \citenamefont {Liu}, \citenamefont {He}, \citenamefont {Li}, \citenamefont {Liu}, \citenamefont {Li}, \citenamefont {Wang}, \citenamefont {Liu},\ and\ \citenamefont {Zhan}}]{fu2022high}%
  \BibitemOpen
  \bibfield  {author} {\bibinfo {author} {\bibfnamefont {Zhuo}\ \bibnamefont {Fu}}, \bibinfo {author} {\bibfnamefont {Peng}\ \bibnamefont {Xu}}, \bibinfo {author} {\bibfnamefont {Yuan}\ \bibnamefont {Sun}}, \bibinfo {author} {\bibfnamefont {Yang-Yang}\ \bibnamefont {Liu}}, \bibinfo {author} {\bibfnamefont {Xiao-Dong}\ \bibnamefont {He}}, \bibinfo {author} {\bibfnamefont {Xiao}\ \bibnamefont {Li}}, \bibinfo {author} {\bibfnamefont {Min}\ \bibnamefont {Liu}}, \bibinfo {author} {\bibfnamefont {Run-Bing}\ \bibnamefont {Li}}, \bibinfo {author} {\bibfnamefont {Jin}\ \bibnamefont {Wang}}, \bibinfo {author} {\bibfnamefont {Liang}\ \bibnamefont {Liu}}, \ and\ \bibinfo {author} {\bibfnamefont {Ming-Sheng}\ \bibnamefont {Zhan}},\ }\bibfield  {title} {\enquote {\bibinfo {title} {High-fidelity entanglement of neutral atoms via a rydberg-mediated single-modulated-pulse controlled-phase gate},}\ }\href {\doibase 10.1103/PhysRevA.105.042430} {\bibfield  {journal} {\bibinfo  {journal} {Phys. Rev. A}\ }\textbf {\bibinfo
  {volume} {105}},\ \bibinfo {pages} {042430} (\bibinfo {year} {2022})}\BibitemShut {NoStop}%
\bibitem [{\citenamefont {Li}\ \emph {et~al.}(2022{\natexlab{c}})\citenamefont {Li}, \citenamefont {Shao},\ and\ \citenamefont {Li}}]{li2022single}%
  \BibitemOpen
  \bibfield  {author} {\bibinfo {author} {\bibfnamefont {X.~X.}\ \bibnamefont {Li}}, \bibinfo {author} {\bibfnamefont {X.~Q.}\ \bibnamefont {Shao}}, \ and\ \bibinfo {author} {\bibfnamefont {Weibin}\ \bibnamefont {Li}},\ }\bibfield  {title} {\enquote {\bibinfo {title} {Single temporal-pulse-modulated parameterized controlled-phase gate for rydberg atoms},}\ }\href {\doibase 10.1103/PhysRevApplied.18.044042} {\bibfield  {journal} {\bibinfo  {journal} {Phys. Rev. Appl.}\ }\textbf {\bibinfo {volume} {18}},\ \bibinfo {pages} {044042} (\bibinfo {year} {2022}{\natexlab{c}})}\BibitemShut {NoStop}%
\bibitem [{\citenamefont {Otterbach}\ \emph {et~al.}(2017)\citenamefont {Otterbach}, \citenamefont {Manenti}, \citenamefont {Alidoust}, \citenamefont {Bestwick}, \citenamefont {Block}, \citenamefont {Bloom}, \citenamefont {Caldwell}, \citenamefont {Didier}, \citenamefont {Fried}, \citenamefont {Hong}, \citenamefont {Karalekas}, \citenamefont {Osborn}, \citenamefont {Papageorge}, \citenamefont {Peterson}, \citenamefont {Prawiroatmodjo}, \citenamefont {Rubin}, \citenamefont {Ryan}, \citenamefont {Scarabelli}, \citenamefont {Scheer}, \citenamefont {Sete}, \citenamefont {Sivarajah}, \citenamefont {Smith}, \citenamefont {Staley}, \citenamefont {Tezak}, \citenamefont {Zeng}, \citenamefont {Hudson}, \citenamefont {Johnson}, \citenamefont {Reagor}, \citenamefont {da~Silva},\ and\ \citenamefont {Rigetti}}]{otterbach2017unsupervised}%
  \BibitemOpen
  \bibfield  {author} {\bibinfo {author} {\bibfnamefont {J.~S.}\ \bibnamefont {Otterbach}}, \bibinfo {author} {\bibfnamefont {R.}~\bibnamefont {Manenti}}, \bibinfo {author} {\bibfnamefont {N.}~\bibnamefont {Alidoust}}, \bibinfo {author} {\bibfnamefont {A.}~\bibnamefont {Bestwick}}, \bibinfo {author} {\bibfnamefont {M.}~\bibnamefont {Block}}, \bibinfo {author} {\bibfnamefont {B.}~\bibnamefont {Bloom}}, \bibinfo {author} {\bibfnamefont {S.}~\bibnamefont {Caldwell}}, \bibinfo {author} {\bibfnamefont {N.}~\bibnamefont {Didier}}, \bibinfo {author} {\bibfnamefont {E.~Schuyler}\ \bibnamefont {Fried}}, \bibinfo {author} {\bibfnamefont {S.}~\bibnamefont {Hong}}, \bibinfo {author} {\bibfnamefont {P.}~\bibnamefont {Karalekas}}, \bibinfo {author} {\bibfnamefont {C.~B.}\ \bibnamefont {Osborn}}, \bibinfo {author} {\bibfnamefont {A.}~\bibnamefont {Papageorge}}, \bibinfo {author} {\bibfnamefont {E.~C.}\ \bibnamefont {Peterson}}, \bibinfo {author} {\bibfnamefont {G.}~\bibnamefont {Prawiroatmodjo}}, \bibinfo {author}
  {\bibfnamefont {N.}~\bibnamefont {Rubin}}, \bibinfo {author} {\bibfnamefont {Colm~A.}\ \bibnamefont {Ryan}}, \bibinfo {author} {\bibfnamefont {D.}~\bibnamefont {Scarabelli}}, \bibinfo {author} {\bibfnamefont {M.}~\bibnamefont {Scheer}}, \bibinfo {author} {\bibfnamefont {E.~A.}\ \bibnamefont {Sete}}, \bibinfo {author} {\bibfnamefont {P.}~\bibnamefont {Sivarajah}}, \bibinfo {author} {\bibfnamefont {Robert~S.}\ \bibnamefont {Smith}}, \bibinfo {author} {\bibfnamefont {A.}~\bibnamefont {Staley}}, \bibinfo {author} {\bibfnamefont {N.}~\bibnamefont {Tezak}}, \bibinfo {author} {\bibfnamefont {W.~J.}\ \bibnamefont {Zeng}}, \bibinfo {author} {\bibfnamefont {A.}~\bibnamefont {Hudson}}, \bibinfo {author} {\bibfnamefont {Blake~R.}\ \bibnamefont {Johnson}}, \bibinfo {author} {\bibfnamefont {M.}~\bibnamefont {Reagor}}, \bibinfo {author} {\bibfnamefont {M.~P.}\ \bibnamefont {da~Silva}}, \ and\ \bibinfo {author} {\bibfnamefont {C.}~\bibnamefont {Rigetti}},\ }\href {https://arxiv.org/abs/1712.05771} {\enquote {\bibinfo
  {title} {Unsupervised machine learning on a hybrid quantum computer},}\ } (\bibinfo {year} {2017}),\ \Eprint {http://arxiv.org/abs/1712.05771} {arXiv:1712.05771 [quant-ph]} \BibitemShut {NoStop}%
\bibitem [{\citenamefont {Bengtsson}\ \emph {et~al.}(2019)\citenamefont {Bengtsson}, \citenamefont {Vikst{\aa}l}, \citenamefont {Warren}, \citenamefont {Svensson}, \citenamefont {Gu}, \citenamefont {Kockum}, \citenamefont {Krantz}, \citenamefont {Krizan}, \citenamefont {Shiri}, \citenamefont {Svensson} \emph {et~al.}}]{bengtsson2019quantum}%
  \BibitemOpen
  \bibfield  {author} {\bibinfo {author} {\bibfnamefont {Andreas}\ \bibnamefont {Bengtsson}}, \bibinfo {author} {\bibfnamefont {Pontus}\ \bibnamefont {Vikst{\aa}l}}, \bibinfo {author} {\bibfnamefont {Christopher}\ \bibnamefont {Warren}}, \bibinfo {author} {\bibfnamefont {Marika}\ \bibnamefont {Svensson}}, \bibinfo {author} {\bibfnamefont {Xiu}\ \bibnamefont {Gu}}, \bibinfo {author} {\bibfnamefont {Anton~Frisk}\ \bibnamefont {Kockum}}, \bibinfo {author} {\bibfnamefont {Philip}\ \bibnamefont {Krantz}}, \bibinfo {author} {\bibfnamefont {Christian}\ \bibnamefont {Krizan}}, \bibinfo {author} {\bibfnamefont {Daryoush}\ \bibnamefont {Shiri}}, \bibinfo {author} {\bibfnamefont {Ida-Maria}\ \bibnamefont {Svensson}},  \emph {et~al.},\ }\bibfield  {title} {\enquote {\bibinfo {title} {Quantum approximate optimization of the exact-cover problem on a superconducting quantum processor},}\ }\href {https://arxiv.org/abs/1912.10495} {\bibfield  {journal} {\bibinfo  {journal} {arXiv preprint arXiv:1912.10495}\ } (\bibinfo {year}
  {2019})}\BibitemShut {NoStop}%
\bibitem [{\citenamefont {{\v{S}}ibali{\'c}}\ \emph {et~al.}(2017)\citenamefont {{\v{S}}ibali{\'c}}, \citenamefont {Pritchard}, \citenamefont {Adams},\ and\ \citenamefont {Weatherill}}]{vsibalic2017arc}%
  \BibitemOpen
  \bibfield  {author} {\bibinfo {author} {\bibfnamefont {Nikola}\ \bibnamefont {{\v{S}}ibali{\'c}}}, \bibinfo {author} {\bibfnamefont {Jonathan~D}\ \bibnamefont {Pritchard}}, \bibinfo {author} {\bibfnamefont {Charles~S}\ \bibnamefont {Adams}}, \ and\ \bibinfo {author} {\bibfnamefont {Kevin~J}\ \bibnamefont {Weatherill}},\ }\bibfield  {title} {\enquote {\bibinfo {title} {Arc: An open-source library for calculating properties of alkali rydberg atoms},}\ }\href {https://doi.org/10.1016/j.cpc.2017.06.015} {\bibfield  {journal} {\bibinfo  {journal} {Computer Physics Communications}\ }\textbf {\bibinfo {volume} {220}},\ \bibinfo {pages} {319--331} (\bibinfo {year} {2017})}\BibitemShut {NoStop}%
\bibitem [{\citenamefont {de~L\'es\'eleuc}\ \emph {et~al.}(2018)\citenamefont {de~L\'es\'eleuc}, \citenamefont {Barredo}, \citenamefont {Lienhard}, \citenamefont {Browaeys},\ and\ \citenamefont {Lahaye}}]{de2018analysis}%
  \BibitemOpen
  \bibfield  {author} {\bibinfo {author} {\bibfnamefont {Sylvain}\ \bibnamefont {de~L\'es\'eleuc}}, \bibinfo {author} {\bibfnamefont {Daniel}\ \bibnamefont {Barredo}}, \bibinfo {author} {\bibfnamefont {Vincent}\ \bibnamefont {Lienhard}}, \bibinfo {author} {\bibfnamefont {Antoine}\ \bibnamefont {Browaeys}}, \ and\ \bibinfo {author} {\bibfnamefont {Thierry}\ \bibnamefont {Lahaye}},\ }\bibfield  {title} {\enquote {\bibinfo {title} {Analysis of imperfections in the coherent optical excitation of single atoms to rydberg states},}\ }\href {\doibase 10.1103/PhysRevA.97.053803} {\bibfield  {journal} {\bibinfo  {journal} {Phys. Rev. A}\ }\textbf {\bibinfo {volume} {97}},\ \bibinfo {pages} {053803} (\bibinfo {year} {2018})}\BibitemShut {NoStop}%
\bibitem [{\citenamefont {Lee}\ \emph {et~al.}(2019)\citenamefont {Lee}, \citenamefont {Kim}, \citenamefont {Jo}, \citenamefont {Song},\ and\ \citenamefont {Ahn}}]{lee2019coherent}%
  \BibitemOpen
  \bibfield  {author} {\bibinfo {author} {\bibfnamefont {Woojun}\ \bibnamefont {Lee}}, \bibinfo {author} {\bibfnamefont {Minhyuk}\ \bibnamefont {Kim}}, \bibinfo {author} {\bibfnamefont {Hanlae}\ \bibnamefont {Jo}}, \bibinfo {author} {\bibfnamefont {Yunheung}\ \bibnamefont {Song}}, \ and\ \bibinfo {author} {\bibfnamefont {Jaewook}\ \bibnamefont {Ahn}},\ }\bibfield  {title} {\enquote {\bibinfo {title} {Coherent and dissipative dynamics of entangled few-body systems of rydberg atoms},}\ }\href {\doibase 10.1103/PhysRevA.99.043404} {\bibfield  {journal} {\bibinfo  {journal} {Phys. Rev. A}\ }\textbf {\bibinfo {volume} {99}},\ \bibinfo {pages} {043404} (\bibinfo {year} {2019})}\BibitemShut {NoStop}%
\bibitem [{\citenamefont {Tamura}\ \emph {et~al.}(2020)\citenamefont {Tamura}, \citenamefont {Yamakoshi},\ and\ \citenamefont {Nakagawa}}]{tamura2020analysis}%
  \BibitemOpen
  \bibfield  {author} {\bibinfo {author} {\bibfnamefont {Hikaru}\ \bibnamefont {Tamura}}, \bibinfo {author} {\bibfnamefont {Tomotake}\ \bibnamefont {Yamakoshi}}, \ and\ \bibinfo {author} {\bibfnamefont {Ken'ichi}\ \bibnamefont {Nakagawa}},\ }\bibfield  {title} {\enquote {\bibinfo {title} {Analysis of coherent dynamics of a rydberg-atom quantum simulator},}\ }\href {\doibase 10.1103/PhysRevA.101.043421} {\bibfield  {journal} {\bibinfo  {journal} {Phys. Rev. A}\ }\textbf {\bibinfo {volume} {101}},\ \bibinfo {pages} {043421} (\bibinfo {year} {2020})}\BibitemShut {NoStop}%
\bibitem [{\citenamefont {Madjarov}\ \emph {et~al.}(2020)\citenamefont {Madjarov}, \citenamefont {Covey}, \citenamefont {Shaw}, \citenamefont {Choi}, \citenamefont {Kale}, \citenamefont {Cooper}, \citenamefont {Pichler}, \citenamefont {Schkolnik}, \citenamefont {Williams},\ and\ \citenamefont {Endres}}]{madjarov2020high}%
  \BibitemOpen
  \bibfield  {author} {\bibinfo {author} {\bibfnamefont {Ivaylo~S}\ \bibnamefont {Madjarov}}, \bibinfo {author} {\bibfnamefont {Jacob~P}\ \bibnamefont {Covey}}, \bibinfo {author} {\bibfnamefont {Adam~L}\ \bibnamefont {Shaw}}, \bibinfo {author} {\bibfnamefont {Joonhee}\ \bibnamefont {Choi}}, \bibinfo {author} {\bibfnamefont {Anant}\ \bibnamefont {Kale}}, \bibinfo {author} {\bibfnamefont {Alexandre}\ \bibnamefont {Cooper}}, \bibinfo {author} {\bibfnamefont {Hannes}\ \bibnamefont {Pichler}}, \bibinfo {author} {\bibfnamefont {Vladimir}\ \bibnamefont {Schkolnik}}, \bibinfo {author} {\bibfnamefont {Jason~R}\ \bibnamefont {Williams}}, \ and\ \bibinfo {author} {\bibfnamefont {Manuel}\ \bibnamefont {Endres}},\ }\bibfield  {title} {\enquote {\bibinfo {title} {High-fidelity entanglement and detection of alkaline-earth rydberg atoms},}\ }\href {https://doi.org/10.1038/s41567-020-0903-z} {\bibfield  {journal} {\bibinfo  {journal} {Nature Physics}\ }\textbf {\bibinfo {volume} {16}},\ \bibinfo {pages} {857--861} (\bibinfo
  {year} {2020})}\BibitemShut {NoStop}%
\bibitem [{\citenamefont {Ivanov}\ and\ \citenamefont {Vitanov}(2015)}]{PhysRevA.92.022333}%
  \BibitemOpen
  \bibfield  {author} {\bibinfo {author} {\bibfnamefont {Svetoslav~S.}\ \bibnamefont {Ivanov}}\ and\ \bibinfo {author} {\bibfnamefont {Nikolay~V.}\ \bibnamefont {Vitanov}},\ }\bibfield  {title} {\enquote {\bibinfo {title} {Composite two-qubit gates},}\ }\href {\doibase 10.1103/PhysRevA.92.022333} {\bibfield  {journal} {\bibinfo  {journal} {Phys. Rev. A}\ }\textbf {\bibinfo {volume} {92}},\ \bibinfo {pages} {022333} (\bibinfo {year} {2015})}\BibitemShut {NoStop}%
\end{thebibliography}%

\end{document}